\documentclass[journal,final]{IEEEtran}

\usepackage{amsmath}
\usepackage{amssymb}
\usepackage{amsfonts}
\usepackage{amsthm}

\usepackage{txfonts}
\usepackage[mathcal]{euscript}

\usepackage{subfigure}
\usepackage{tikz}
\usepackage{xcolor}
\definecolor{halfgray}{gray}{0.55}
\definecolor{webgreen}{rgb}{0,.5,0}
\definecolor{webbrown}{rgb}{.6,0,0}
\definecolor{BrightViolet}{rgb}{0.5,0.2,0.8}
\definecolor{Maroon}{cmyk}{0, 0.87, 0.68, 0.32}
\definecolor{RoyalBlue}{cmyk}{1, 0.50, 0, 0}
\definecolor{Black}{cmyk}{0, 0, 0, 0}

\usepackage[dots]{12many}
\usepackage[smaller]{acronym}
\usepackage{balance}
\usepackage{paralist}

\newcommand{\dis}{\displaystyle}
\newcommand{\mb}{\mathbf}
\newcommand{\txs}{\textstyle}
\newcommand{\insum}{\sum\nolimits}
\newcommand{\inprod}{\prod\nolimits}

\newcommand{\dd}{\,d}
\newcommand{\eps}{\varepsilon}

\newcommand{\pd}{\partial}

\newcommand{\parl}{\varparallel}
\newcommand{\simplex}{\Delta}

\newcommand{\set}{\mathcal{S}}
\newcommand{\play}{\mathcal{K}}
\newcommand{\act}{\mathcal{A}}

\newcommand{\strat}{\Delta}
\newcommand{\graph}{\mathcal{G}}
\newcommand{\edges}{\mathcal{E}}

\newcommand{\cardlinks}{Q}

\newcommand{\game}{\mathfrak{G}}

\newcommand{\cone}{T^{c}}

\newcommand{\dkl}{D_{\textup{KL}}}

\newcommand{\midd}{\,\|\,}
\newcommand{\ddt}{\frac{d}{dt}}

\newcommand{\R}{{\mathbb R}}

\DeclareMathOperator{\exclude}{\setminus}

\DeclareMathOperator{\ex}{\mathbf{E}}

\DeclareMathOperator{\bigoh}{\mathcal O}

\DeclareMathOperator{\ind}{ind}

\DeclareMathOperator{\Ei}{Ei}
\DeclareMathOperator{\supp}{supp}

\DeclareMathOperator{\card}{card}
\DeclareMathOperator{\sre}{\sf SRE}
\DeclareMathOperator{\sinr}{\sf sinr}
\DeclareMathOperator{\eql}{\sf EQL}

\newlength{\wideword}
\newlength{\leftover}
\theoremstyle{plain}
\newtheorem{theorem}{Theorem}
\newtheorem{corollary}{Corollary}
\newtheorem*{corollary*}{Corollary}
\newtheorem{lemma}{Lemma}
\newtheorem{proposition}{Proposition}

\theoremstyle{definition}
\newtheorem{definition}{Definition}
\newtheorem*{definition*}{Definition}

\theoremstyle{remark}
\newtheorem{remark}{Remark}
\newtheorem*{remark*}{Remark}

\begin{document}



\newacro{IC}{interference channel}
\newacro{WLAN}{wireless local area network}
\newacro{DSL}{digital subscriber line}
\newacro{MC-CDMA}{multi-carrier code division multiple access}
\newacro{SUD}{single-user decoding}
\newacro{MAC}{multiple access channel}
\newacro{MIMO}{multiple-input multiple-output}
\newacro{PMAC}{parallel multiple access channel}
\newacro{OFDM}{orthogonal frequency-division multiplexing}
\newacro{SIC}{successive interference cancellation}
\newacro{SNR}{signal to noise ratio}
\acused{SNR}
\newacro{CDF}{cumulative distribution function}
\acused{CDF}
\newacro{SINR}{signal to interference-plus-noise ratio}
\newacro{BR}{best-response}
\newacro{RL}{reinforcement learning}
\newacro{SDE}{stochastic differential equation}
\newacro{SRE}{sum-rate efficiency}
\acused{SRE}
\newacro{EQL}{equilibration level}
\newacro{KKT}{Karush-Kuhn-Tucker}
\newacro{ODE}{ordinary differential equation}


\title{Distributed Learning Policies for\\Power Allocation in Multiple Access Channels}

\date{\today}
\author{
Panayotis Mertikopoulos,
Elena~V.~Belmega,~\IEEEmembership{Member,~IEEE},
Aris~L.~Moustakas,~\IEEEmembership{Senior Member,~IEEE},
and Samson Lasaulce,~\IEEEmembership{Member,~IEEE}
\thanks{
This work was supported in part by the Greek GSRT ``Kapodistrias'' project No. 70/3/8831, the GIS ``Sciences de la D\'ecision'' (\'Ecole Polytechnique\textendash\ ENSAE\textendash\ HEC), and the P\^ole de Recherche en Economie et Gestion (UMR 7176).
The work was initiated when E.~V.~Belmega was visiting the University of Athens during the summer of 2010 with the support of the L'Or\'eal program ``For young women doctoral candidates in science 2009''.
Part of this work was presented in the 5th International Conference on Performance Evaluation Methodologies and Tools (\textsc{ValueTools '11}), Paris, May 2011.}

\thanks{P.~Mertikopoulos is with the Centre National de la Recherche Scientifique (CNRS) and the Laboratoire d'Informatique de Grenoble, France; at the time of submission, he was with the Economics Department of \'Ecole Polytechnique, Paris, France (e-mail: panayotis.mertikopoulos@imag.fr).}

\thanks{E.V. Belmega is with ETIS/ENSEA \textendash\ Universit\'e de Cergy-Pontoise \textendash\ CNRS, Cergy-Pontoise, France; at submission, she was with the Alcatel-Lucent Chair on Flexible Radio in Sup\'elec, France (email: belmega@ensea.fr).}

\thanks{A.~L.~Moustakas is with the Physics Department of the University of Athens, Greece (e-mail: arislm@phys.uoa.gr).}

\thanks{S.~Lasaulce is with the Laboratoire des Signaux et Syst\`emes, Sup\'elec, Gif-sur-Yvette, France (e-mail: samson.lasaulce@lss.supelec.fr.}


}


\maketitle


\begin{abstract}
We analyze the power allocation problem for orthogonal multiple access channels by means of a non-cooperative potential game in which each user distributes his power over the channels available to him. When the channels are static, we show that this game possesses a unique equilibrium; moreover, if the network's users follow a distributed learning scheme based on the replicator dynamics of evolutionary game theory, then they converge to equilibrium exponentially fast. On the other hand, if the channels fluctuate stochastically over time, the associated game still admits a unique equilibrium, but the learning process is not deterministic; just the same, by employing the theory of stochastic approximation, we find that users still converge to equilibrium.

Our theoretical analysis hinges on a novel result which is of independent interest: in finite-player games which admit a (possibly nonlinear) convex potential, the replicator dynamics converge to an $\eps$-neighborhood of an equilibrium in time $\bigoh(\log(1/\eps))$.
\end{abstract}



\section{Introduction}
\label{sec:introduction}

\IEEEPARstart{I}{n} view of the decentralized nature of future and emerging wireless networks, non-cooperative game theory has become an important tool to analyze distributed problems in networks whose nodes cannot be assumed to adhere to centrally controlled protocols. The main goal has been to develop policies and algorithms that nodes can use to optimize their resources (power, bandwidth, etc.) on their own, so, following \cite{LDA09}, the questions that arise are
\begin{inparaenum}[\itshape a\upshape)]
\item whether there exist ``equilibrial'' policies which are stable against unilateral deviations;
\item whether these (Nash) equilibria are unique; and
\item whether they can be reached by distributed learning algorithms that require only local information.
\end{inparaenum}

Accordingly, an important paradigm which has attracted significant interest in the wireless communications literature concerns the allocation of power over orthogonal communication channels \cite{LDA09,GT06,MPS07}.
From a centralized viewpoint, this is a relatively well-studied subject, especially with respect to optimal power allocation schemes which allow users to reach the boundary of the rate region assuming full channel knowledge and central control \cite{CV93, TH98}. On the other hand, more recent examinations \cite{ABSA02,SPB08-jsac,SPB09-sp} focus on socially stable power allocation policies because, even if the globally optimal, capacity-achieving power profile is known and used, it might be unstable under deviations by selfish users (and thus useless in a decentralized setting).

In this paper, we consider the problem of uplink communication in multi-user networks consisting of several receivers that operate on distinct, non-interfering channels, and we focus on giving definitive answers to points (a)\textendash(c) above, analyzing the equilibrial structure of the problem and its convergence aspects. Despite its apparent simplicity, this \ac{PMAC} model has several relevant applications such as, for instance, in 802.11-based \acp{WLAN} with non-overlapping channels \cite{CHD10,LSdLD08}, distributed soft handoffs in cellular systems \cite{PBLD09}, distributed power allocation in  \acp{DSL} \cite{YGC02}, and, finally, in throughput-maximizing power control in \ac{MC-CDMA} systems \cite{MCPS06}.

Our analysis will focus on the \ac{SUD} scheme where the transmitted signal of each user is decoded separately by the receiver(s) who treat the incoming signal of other users as additive (Gaussian) noise. The main reason for using \ac{SUD} instead of \ac{SIC} is that the former is known to have lower decoding complexity and signalling overhead than the latter \textendash\ a consequence of \ac{SUD} not having to broadcast the decoding order to the transmitters \cite{BelmegaThesis}. As a result, \ac{SIC}-based schemes suffer from scalability issues, especially when there are several receivers and/or the channel is highly time-varying.

In this context, non-cooperative power allocation games for static Gaussian \acp{IC} have been studied in a series of related papers \cite{SPB08-jsac,SPB09-sp}. There, the existence of a Nash equilibrium is a consequence of the convexity properties of the users' achievable rates and follows from the general theory of \cite{Ro65}. In fact, under suitable (but stringent) conditions on the channel matrices, it was shown that this equilibrium is unique and that iterative water-filling algorithms converge to it.

Formally, the static \ac{PMAC} is a special case of this \ac{IC} framework, but the conditional analysis of \cite{SPB08-jsac,SPB09-sp} almost always fails for static \ac{PMAC} models. Thus, although the (global) capacity region of this channel is well-understood \cite{CV93,TH98,GT06}, the channel's distributed version remains unresolved. A first attempt to remedy this was carried out in \cite{PBLD09} where it was shown that an associated power allocation game admits an exact potential function \cite{MS96} whose minimum corresponds to the game's Nash equilibria. However, this potential is \emph{not} strictly convex, so Nash equilibrium uniqueness might fail along with the uniqueness conditions of \cite{SPB08-jsac,SPB09-sp}. Rather surprisingly, it turns out that this is \emph{not} the case:
even though these conditions do not hold in the static \ac{PMAC} context, the Nash equilibrium of the static \ac{PMAC} game \emph{is} unique (Theorem~\ref{thm:uniqueness}).

In itself, uniqueness allows us to characterize the system's behavior at equilibrium, but it does not provide a way of actually getting there. Regarding such convergence issues, the authors of \cite{ABSA02} considered a single channel with pricing and exhibited power control algorithms which converge to equilibrium under ``mild-interference'' conditions. Similarly, one of the main results of \cite{YGC02} was to show that if the transmitters know the local channel state and the overall inteference-plus-noise covariance matrix, then, subject to similar ``mild-interference'' conditions, iterative water-filling converges to the equilibrium set of the game (a result which was then enhanced in \cite{YRBC04} by dropping this condition for a modified water-filling scheme).

Instead of taking a water-filling approach, we present a simpler learning scheme based on the replicator dynamics of evolutionary game theory \cite{TJ78} which involves the same (often less) information from the side of the players, and which does not require them to solve a nonlinear water-filling problem. Dynamics of this type have been studied extensively in finite \cite{FL98} and continuous population games \cite{Sandholm11}, but, in \emph{nonlinear games} (such as the one we have here), their properties are not as well understood. Nonetheless, by taking a modified version of the users' utilities, we show that the replicator dynamics converge to an $\eps$-neighborhood of the game's (a.s.) unique equilibrium exponentially fast, i.e. in time $\bigoh(\log(1/\eps))$ (Theorem \ref{thm:convergence}).

In the context of fading however, the static game and the corresponding replicator dynamics lose much of their relevance because variations due to fading open the door to stochasticity. To account for this randomness, we study both fast and block-fading models. Using techniques from the theory of stochastic approximation \cite{Borkar08}, we show that by properly adjusting their learning scheme, users converge to the (unique) equilibrium of an averaged game whose payoff functions correspond to the users' achievable ergodic rates (Theorems \ref{thm:stochconvergence} and \ref{thm:ergconvergence}).

Our convergence analysis is based on a novel game-theoretic result which is of independent interest: {\em in games which admit a (star-)convex potential function, the replicator dynamics converge to (the game's unique) equilibrium at an exponential rate} (Theorem \ref{thm:convspeed}). To the best of our knowledge, this is the fastest convergence rate that has been established for the replicator dynamics in the current state of the art \cite{Sandholm11}.

\paragraph*{Notational Conventions}
If $\R^{\set}$ is the vector space spanned by the set $\set= \{s_{\alpha}\}_{\alpha=1}^{S}$ and $\{e_{\alpha}\}_{\alpha=1}^{S}$ denotes its canonical basis, we will use $\alpha$ to refer interchangeably to either $s_{\alpha}$ or $e_{\alpha}$, and we will identify the set $\simplex(\set)$ of probability measures on $\set$ with the standard $(S-1)$-dimensional simplex of $\R^{\set}$: $\simplex(\set) \equiv \{x\in\R^{\set}: \insum_{\alpha} x_{\alpha} =1 \text{ and } x_{\alpha}\geq0\}$. Finally, we will employ Latin indices for players ($k,\ell,\dotso$), while reserving Greek ones ($\alpha,\beta,\dotso$) for their (``pure'') strategies; and when summing over $\alpha\in\act_{k}$, we will simply write $\sum^{k}_{\alpha} \equiv \sum_{\alpha\in\act_{k}}$.

\section{System Model}
\label{sec:model}

Following \cite{PBLD09}, the basic setup of our model is as follows: we consider a finite set $\play=\ito{K}$ of wireless single-antenna transmitters (the {\em players} of the game) who wish to transmit to a group of single-antenna receivers (possibly clustered as a single receiver). Each of these receivers operates on a given channel $\alpha\in\act\equiv\ito{A}$ (assumed to be orthogonal, typically in the frequency domain), and each user $k\in\play$ may transmit over a subset $\act_{k}\subseteq\act$ of these channels (with $A_{k}\equiv \card(\act_{k})\geq2$).

In particular, if $x_{k\alpha}\sim\mathcal{CN}(0,p_{k\alpha})$ is the transmitted message of user $k$ on channel $\alpha\in\act_{k}$ and $h_{k\alpha}$ denotes the respective channel coefficient, then the received signal on channel $\alpha$ will be $y_{\alpha} = \insum_{k} h_{k\alpha} x_{k\alpha} + z_{\alpha}$, where $z_{\alpha} \sim \mathcal{CN}(0,\sigma_{\alpha}^{2})$ denotes the thermal noise. Accordingly, user $k\in\play$ can split his transmitting power among the channels $\alpha\in\act_{k}$ subject to the constraint:
\begin{equation}
\label{eq:powerconstraint}
\txs
\insum^{k}_{\alpha} p_{k\alpha} \leq P_{k},
\end{equation}
where $p_{k\alpha} = \ex\big[|x_{k\alpha}|^{2}\big]$ represents the power with which the user transmits on channel $\alpha$, and $P_{k}$ is his maximum power. As a result, the {\em power allocation} of user $k$ will be given by the point $p_{k} = \sum^{k}_{\alpha} p_{k\alpha} e_{k\alpha}\in\R^{\act_{k}}$, and, analogously, the {\em power profile} which collects all users' power allocations will be re\-pre\-sented by $p = (p_{1},\dotsc,p_{K})\in\prod_{k}\R^{\act_{k}}=\R^{\cardlinks}$ with $\cardlinks \equiv \insum_{k} A_{k}$.

In this context, our performance metric will be the users' achievable transmission rates, which depend on their \ac{SINR}:
\begin{equation}
\label{eq:sinr}
\sinr_{k\alpha}(p) = \frac{g_{k\alpha} p_{k\alpha}}{\sigma_{\alpha}^{2} + \insum_{\ell\neq k} g_{\ell\alpha} p_{\ell\alpha}},
\end{equation}
with $g_{k\alpha}=|h_{k\alpha}|^{2}$ denoting the channel gain coefficient of user $k$ in channel $\alpha\in\act_{k}$. Clearly, the users' achievable rates will depend on their power allocation policies through their \ac{SINR}, but the exact dependence hinges on the time-variability of the channel gain coefficients $g_{k\alpha}$.

At one end, we will study \emph{static} channels, i.e. channels whose coherence time is much larger than both the self-decodable block duration and the power updating period. At the other extreme, we will also consider \emph{fast-fading} channels where the coherence time is much shorter than those characteristic times; here, what matters is the ergodic value of the \ac{SINR} and the corresponding rate. Finally, we will also analyze the more interesting intermediate case, where the coherence time is greater than the block length but comparable to the update time \textendash\ hence allowing blocks to be decoded using the instantaneous channel values in (\ref{eq:sinr}), but also introducing stochasticity in the game.

\subsection{Static Channels}
\label{subsec:static.game}

We will start with the case of static channels employing the \acf{SUD}. In this case, the spectral efficiency of user $k$ in the power profile $p$ will be given by \cite{PBLD09,ABSA02}:
\begin{equation}
\label{eq:payoff}
\txs
u_{k}(p)
= \insum_{\alpha} b_{\alpha} \log\Big(1 + \sinr_{k\alpha}(p)\Big)
\end{equation}
where $b_{\alpha}>0$ represents the bandwidth of channel $\alpha\in\act_{k}$ and the channel gains $g_{k\alpha}$ are drawn once and for all from a continuous distribution on $[0,\infty)$ at the outset of the game and remain fixed for the duration of the transmission \cite{SPB08-jsac,PBLD09}.
Then, to maximize their spectral efficiency (\ref{eq:payoff}), users saturate (\ref{eq:powerconstraint}) by transmitting at the highest possible power \cite{PBLD09}, so we are led to the \emph{static} \ac{PMAC} game $\game\equiv\game\big(\play,\{\strat_{k}\},\{u_{k}\}\big)$ where:
\begin{enumerate}
\item The {\em players} of $\game$ are the transmitters $\play = \ito{K}$.
\item The {\em strategy space} of player $k$ is the (scaled) simplex $\strat_{k} \equiv P_{k}\,\simplex(\act_{k}) = \left\{p_{k}\in\R^{\act_{k}}: p_{k\alpha}\geq 0 \text{ and }\sum_{\alpha} p_{k\alpha}=P_{k}\right\}$ of power allocation vectors; the game's space of strategy profiles $p=(p_{1},\dotsc,p_{K})$ will then be $\strat\equiv\prod_{k}\strat_{k}$.
\item The players' {\em payoffs} (or {\em utilities}) are given by the spectral efficiencies $u_{k}\colon\strat\to\R$ of (\ref{eq:payoff}).
\end{enumerate}

Of course, the game $\game$ defined in this way does not adhere to the original normal form of Nash in the sense that
\begin{inparaenum}[\itshape a\upshape)]
\item players are not mixing probabilities over a finite set of possible actions, and
\item even though the players' strategy spaces are simplices, their payoffs are not (multi)linear.
\end{inparaenum}
On the other hand, with $u_{k}$ being concave in $p_{k}$, it is easy to see that $\game$ is itself {\em concave} in the sense of Rosen \cite{Ro65}. Also, as was shown in \cite{PBLD09}, $\game$ possesses an \emph{exact potential} \cite{MS96}, i.e. a function $\Phi\colon\strat\to\R$ such that:
\begin{equation}
\label{eq:potentialdef}
u_{k}(p_{-k};p_{k}') - u_{k}(p_{-k};p_{k}) = \Phi(p_{-k};p_{k}) - \Phi(p_{-k};p_{k}'),
\end{equation}
for all users $k\in\play$, and for all power allocations $p_{k},p_{k}'\in\strat_{k}$ of user $k$ and $p_{-k}\in\strat_{-k}\equiv\prod_{\ell\neq k}\strat_{\ell}$ of $k$'s opponents;
in fact, the analysis of \cite{PBLD09} also provides the expression:
\begin{equation}
\label{eq:potential}
\txs
\Phi(p) = -\insum_{\alpha} b_{\alpha}
\log\left(1 + \insum_{k} g_{k\alpha} p_{k\alpha}\big/ \sigma_{\alpha}^{2}\right).
\end{equation}

\subsection{Fading Channels}
\label{subsec:fading.game}

As discussed above, the non-static models that we will examine are block-fading and fast-fading channels.

\subsubsection{Block-fading channels}

In this case, the coefficients $g_{k\alpha} \equiv g_{k\alpha}(t)$ remain constant over an entire transmission block, so, assuming the transmitter knows (\ref{eq:sinr}) for each block through feedback, the users' utilities are still given by (\ref{eq:payoff}), with different gains $g_{k\alpha}$ at each self-decodable block.\footnote{Note that we do not assume any delay constraints at the receiver \cite{OSW94}; in this way, reliable information (a la Shannon) can be transmitted over each block.} As such, (\ref{eq:potential}) is still a potential for the (now evolving) game $\game(t)$, the only difference being that $\Phi$ will evolve over time following the channels and the game.

\subsubsection{Fast-fading channels}

In this regime, the coefficients $h_{k\alpha}\equiv h_{k\alpha}(t)$ evolve ergodically at a rate which is much faster than the characteristic length of a transmission block, so the ``instantaneous'' utilities (\ref{eq:payoff}) lose their relevance. Instead, and assuming for simplicity that users saturate their power constraints, their utilities will be given by the \emph{ergodic rates} of \cite{GV97}:
\begin{equation}
\label{eq:ergpayoff}
\txs
\overline u_{k}(p) = \insum_{\alpha} b_{\alpha} \ex_{g}
\left[\log
\left(1+\sinr_{k\alpha}(p)\right)
\right].
\end{equation}
We thus obtain the {\em ergodic game} $\overline\game\equiv\big(\play,\{\strat_{k}\},\{\overline u_{k}\}\big)$, which has the same strategic structure as its static counterpart $\game$ but payoffs given by (\ref{eq:ergpayoff}) instead of (\ref{eq:payoff}).

In fact, as in the static case, $\overline\game$ admits the exact potential:
\begin{equation}
\label{eq:ergpotential}
\txs
\overline\Phi(p)
\equiv \ex_{g}[\Phi(p)]
=-\insum_{\alpha} b_{\alpha} \ex_{g}\big[
\log\big(1+ \insum_{k} g_{k\alpha} p_{k\alpha}\big/\sigma_{\alpha}^{2}\big)
\big],
\end{equation}
whose form depends on the law of the $g_{k\alpha}$. Thus, with $h_{k\alpha}\sim\mathcal{CN}(0,\sqrt{\gamma_{k\alpha}})$, $\gamma_{k\alpha}\geq0$, the coefficients $g_{k\alpha} = |h_{k\alpha}|^{2}$ will be $\chi^{2}$-distributed, and the calculations of \cite[eq. (11)]{MS03} yield:

\begin{proposition}
\label{prop:gaussianpotential}
In i.i.d. Gaussian fast-fading channels with $h_{k\alpha}\sim\mathcal{CN}(0,\sqrt{\gamma_{k\alpha}})$, the ergodic potential $\overline\Phi$ is:
\begin{equation}
\label{eq:Gaussianpotential}
\overline\Phi (p)=
-\insum_{k,\alpha} b_{\alpha} \zeta(r_{k\alpha}^{-1})
\inprod_{\ell\neq k}(1-r_{\ell\alpha}/r_{k\alpha})^{-1}
\end{equation}
where $r_{k\alpha} \!=\! \gamma_{k\alpha}p_{k\alpha}\big/\sigma_{\alpha}^{2}$ and $\zeta(x) \!\equiv \!\int_{0}^{\infty}\!(x+t)^{-1} e^{-t}  \dd t \!=\! - e^{x} \Ei(-x)$.
\end{proposition}

This proposition will be crucial in the numerical calculations of Section~\ref{sec:numerics}. For posterity, we only note here that (\ref{eq:Gaussianpotential}) implies that $\overline\Phi$ is strictly convex \cite{BLDH10}, even though, in general, $\Phi$ is not.

\section{Equilibrium Analysis}
\label{sec:equilibrium}

We begin with the notion of \emph{Nash equilibrium}:
\begin{definition}
A power profile $q\in\strat$ will be a {\em Nash equilibrium} of the game $\game$ (resp. $\overline\game$) when:
\begin{equation}
\label{eq:Nash}
u_{k}(q) \geq u_{k}(q_{-k};q_{k}'),\quad
\text{(resp. $\overline u_{k}(q) \geq \overline u_{k}(q_{-k};q_{k}')$)}
\end{equation}
for all $k\in\play$ and for all $q_{k}'\in\strat_{k}$. In particular, if $q$ satisfies the strict version of (\ref{eq:Nash}) for all $q_{k}'\neq q_{k}$, it will be called \emph{strict}.
\end{definition}

Given that $\game$ (resp. $\overline\game$) admits a convex potential, \emph{its equilibrium set will coincide with the minimum set of $\Phi$} (resp.~$\overline\Phi$) \cite{Neyman97}. As such, the existence of an equilibrium is guaranteed, and this is already important from a practical point of view because learning protocols would never converge otherwise. Our goal in this section will be to show that these equilibria are essentially \emph{unique}, thus ensuring the system's \emph{predictability} \textendash\ a crucial feature for performance evaluation, QoS guarantees, etc.

\subsection{Static Channels}
\label{subseq:static.equilibrium}

With regards to the static potential $\Phi$, it is easy to see that two power profiles $p,p'\in\strat$ will have $\Phi(p) = \Phi(p')$ whenever
\begin{equation}
\label{eq:degconstraint}
\txs
\insum_{k} g_{k\alpha} (p_{k\alpha}' - p_{k\alpha}) = 0\quad
\text{for all $\alpha\in\act$}.
\end{equation}
In that case, $\Phi$ will not be strictly convex, so its minimum set might fail to be a singleton as well. More precisely, if we set $z=p'-p$, we will also have $\sum_{\alpha}^{k} z_{k\alpha}=0$ for all $k\in\play$, so, on account of (\ref{eq:degconstraint}) above, $\Phi$ will not be strictly convex if the following linear system admits a non-zero solution in $z$:
\begin{align}
\label{eq:constraints}
\txs
\insum_{k} g_{k\alpha} z_{k\alpha} = 0,\quad \alpha\in\act;
&&
\txs
\insum_{\alpha} z_{k\alpha} = 0,\quad k\in\play.
\end{align}

Since $z\in\R^{\cardlinks}$, $\cardlinks=\sum_{k}A_{k}$, and the above $A+K$ constraints are independent (a.s.), we see that if $\cardlinks- A - K >0$, then $\Phi$ cannot be strictly convex \textendash\ see \cite{ValueTools11b} for more details. In fact, the quantity $\ind(\game) \equiv \cardlinks - A -K$ will be called the \emph{degeneracy index} of the game $\game$, and the condition $\ind(\game) >0$ means that \emph{if the number of links $Q$ exceeds the number of channels plus transmitters $A+K$, then the game's potential is not strictly convex}.

In typical uplink scenarios of practical interest (e.g. single-receiver \acs{OFDM}), each user can access all channels, so $A_{k}=A$ for all $k$, and, hence $Q = AK > A + K$ (except in small $2\times2$ systems). This implies that degeneracy appears almost always, so in the absence of strict convexity, a promising way to determine whether the \ac{PMAC} game admits a \emph{unique} equilibrium would be to use the conditions of \cite{SPB08-jsac,SPB09-sp,LP06} where one constructs a certain matrix $\mb{S}$ from the channel gain coefficients and tries to show that said matrix has a spectral radius $\rho(\mb{S})<1$. However, as is shown in \cite{ValueTools11b}, this spectral radius exceeds $1$ (a.s.), so the results of \cite{SPB08-jsac,SPB09-sp,LP06} do not apply to our problem. Still, we have:

\begin{theorem}
\label{thm:uniqueness}
The static game $\game$ admits a unique Nash equilibrium (a.s.).
\end{theorem}

\begin{IEEEproof}[Sketch of Proof]
Let $p\in\strat$ and consider the \emph{(multi)graph} $\graph(p) = (\act, \edges(p))$ whose vertices are the network's receivers and whose edge set $\edges(p)$ is the \emph{multiset sum} $\edges(p) = \biguplus_{k} \edges_{k}(p_{k})$ where each $\edges_{k}(p_{k})$ is a star graph on the nodes $\alpha\in\act$ to which $p_{k}$ assigns positive power $p_{k\alpha}>0$ (i.e. all channels to which $k$ transmits with positive power are star-connected and these graphs are su\-per\-imposed for all $k\in\play$). If $p$ is equilibrial, $\graph(p)$ has to be acyclic \cite{ValueTools11b}, so $p$ must lie in the interior of an at most $(A-1)$-dimensional face of $\strat$ (a.s.); our assertion then follows from a dimension-counting argument (see \cite{ValueTools11b} for details).
\end{IEEEproof}

\subsection{Fading Channels}
\label{subsec:ergodic.equilibrium}

\subsubsection{Block-fading channels}

As we discussed in Section \ref{subsec:fading.game}, the time-varying version of (\ref{eq:potential}) which corresponds to block-fading channels $g_{k\alpha} \equiv g_{k\alpha}(t)$ is a potential for the block-fading game $\game(t)$. Accordingly, Theorem \ref{thm:uniqueness} implies:
\begin{corollary}
At each channel realization, the block-fading game $\game(t)$ admits a unique Nash equilibrium (a.s.).
\end{corollary}

\subsubsection{Fast-fading channels}

On the other hand, the averaging effect in the ergodic rates (\ref{eq:ergpayoff}) can be used to show that the ergodic potential $\overline\Phi$ is, in fact, \emph{strictly} convex. This gives:
\begin{theorem}[\cite{BLDH10}]
\label{thm:erguniqueness}
The ergodic game $\overline\game$ admits a unique Nash equilibrium.
\end{theorem}


\section{Learning Dynamics and Convergence to Equilibrium}
\label{sec:convergence}

Although Theorems \ref{thm:uniqueness} and \ref{thm:erguniqueness} guarantee equilibrium uniqueness, it is not at all clear whether users will be able to calculate this equilibrium in decentralized environments where only partial/local information is available at the terminal (e.g., as in distributed or partially distributed cognitive radio networks). Consequently, our goal in this section will be to present a simple distributed learning scheme which allows users to converge to equilibrium, and to determine the speed of this convergence.

This question has attracted considerable interest from the point of view of learning, and two of the most well-studied paradigms are \ac{BR} algorithms and \ac{RL} \cite{BS97,FL98,Sandholm11}. In standard \ac{BR} schemes \cite{Sandholm11}, players are assumed to monitor their opponents' power allocation policies and respond optimally to them (with respect to their individual utilities). Unfortunately (and in addition to the ``perfect monitoring'' requirement), it is quite hard to calculate these best responses in large games, so the applicability of this approach to large decentralized networks is quite limited. To circumvent these limitations (in static channels at least), a promising solution lies in the water-filling approach of \cite{YGC02,SPB08-jsac,SPB09-sp,YRBC04} where users only need their local channel and overall noise-plus-interference covariance matrix. In that case however
\begin{inparaenum}[\itshape a\upshape)]
\item users must solve a non-convex fixed point problem at each step; and
\item convergence is conditional on the interference being low enough (except in \cite{YRBC04}).
\end{inparaenum}
In fact, the conditions of \cite{SPB08-jsac,SPB09-sp} do \emph{not} hold in the \ac{PMAC} case \cite{ValueTools11b}, while the approach of \cite{YGC02} breaks down for large numbers of users \cite{HH10}.

On the other hand, \ac{RL} algorithms (such as regret-matching \cite{HMC00}) rely on the players knowing their (possibly fictitious) payoffs. Thanks to this information (which, however, is often hard to come by), these algorithms enjoy strong convergence properties in potential games. However, such learning algorithms have been designed for discrete action sets, so it is very hard to adapt them to games with continuous action spaces (such as the ones we are considering here).

To overcome these limitations, our starting point will be the replicator dynamics of evolutionary game theory \cite{TJ78,FL98,Sandholm11}. The reinforcement aspect of these dynamics does suffer from the same drawback as most \ac{RL} algorithms (i.e. it applies only to \emph{finite} action sets), but, by exploiting the simplicial structure of the game and its potential, we derive a learning scheme which applies to \emph{continuous} action spaces and which allows users to converge to equilibrium \emph{unconditionally} and \emph{exponentially quickly} (Theorems \ref{thm:convergence} and \ref{thm:ergconvergence}).

\subsection{Static Channels}
\label{subsec:static.convergence}

Since the replicator equation applies to discrete sets (such as $\act_{k}$), a reasonable channel-specific utility would be:
\begin{equation}
\label{eq:nodepayoffs}
u_{k\alpha}(p)
= b_{\alpha}\log\Big(1+\sinr_{k\alpha}(p)\Big)
\end{equation}
which leads to the replicator equation:
\begin{equation}
\label{eq:naiveRD}
\frac{dp_{k\alpha}}{dt}
= p_{k\alpha}(t) \left(u_{k\alpha}(p(t)) - P_{k}^{-1}\insum^{k}_{\beta} p_{k\beta}(t) u_{k\beta}(p(t))\right),
\end{equation}
whose second term ensures that $p(t)\in\strat$ for all $t\geq0$. Unfortunately, the utility $u_{k}$ of eq.~(\ref{eq:payoff}) is \emph{not} a convex combination of the $u_{k\alpha}$, so (\ref{eq:naiveRD}) is not well-behaved w.r.t. the game $\game$ either \textendash\ for instance, Nash equilibria are not stationary.

Instead, given that each user invariably seeks to unilaterally increase his utility, we will consider the \emph{marginal utilities}:
\begin{equation}
\label{eq:derpayoff}
v_{k\alpha}(p)
\equiv \frac{\pd u_{k}}{\pd p_{k\alpha}}
= \frac{b_{\alpha}g_{k\alpha}}{\sigma_{\alpha}^{2} + \insum_{\ell} g_{\ell\alpha} p_{\ell\alpha}}.
\end{equation}
Since player $k$ can calculate $v_{k\alpha}(p)$ by means of $\sinr_{k\alpha}$ and $g_{k\alpha}$ alone (the bandwidths $b_{\alpha}$ are assumed fixed and known), any dynamics based on the $v_{k\alpha}$'s will be inherently distributed (and simpler than solving a water-filling problem to boot). We will thus consider the replicator equation:
\begin{equation}
\label{eq:RD}
\frac{dp_{k\alpha}}{dt}
\txs= p_{k\alpha}(t) \big(v_{k\alpha}(p(t)) - v_{k}(p(t))\big),
\end{equation}
where $v_{k}$ denotes the user average $v_{k}(p) = P_{k}^{-1}\insum^{k}_{\beta} p_{k\beta} v_{k\beta}(p)$.

\begin{remark}[Comparison to other power updating schemes]
The replicator equation (\ref{eq:RD}) is clearly quite unlike the water-filling schemes of \cite{YGC02,SPB08-jsac,SPB09-sp,YRBC04}. Closer in spirit to (\ref{eq:RD}) are the algorithms developed in the 90's with the goal of minimizing transmitting power by comparing the instantaneous \ac{SINR} to a target value and iteratively updating the power proportionally to this difference \cite{FM93,BCP94,Yates95}; still, there is little overlap with these algorithms, both in terms of setup and convergence.
\end{remark}

\begin{remark}
The marginal utilities $v_{k\alpha}$ are similar but {\em not} equal to the \ac{SINR} (\ref{eq:sinr}) of user $k$ at a given channel $\alpha\in\act_{k}$, and they do not coincide with the popular metric of total rate per unit power either \cite{MCPS06,MPS07}. These metrics can all be calculated based on the same feedback (and might appear more appealing than $v_{k\alpha}$), but we shall see that it is precisely the (perhaps unconventional) choice of the marginal utilities $v_{k\alpha}$ that leads to convergence.
\end{remark}

\begin{figure*}
\centering
\subfigure[Convergence of the replicator dynamics in static channels.]{
\label{subfig:portrait}
\includegraphics[width=0.825\columnwidth]{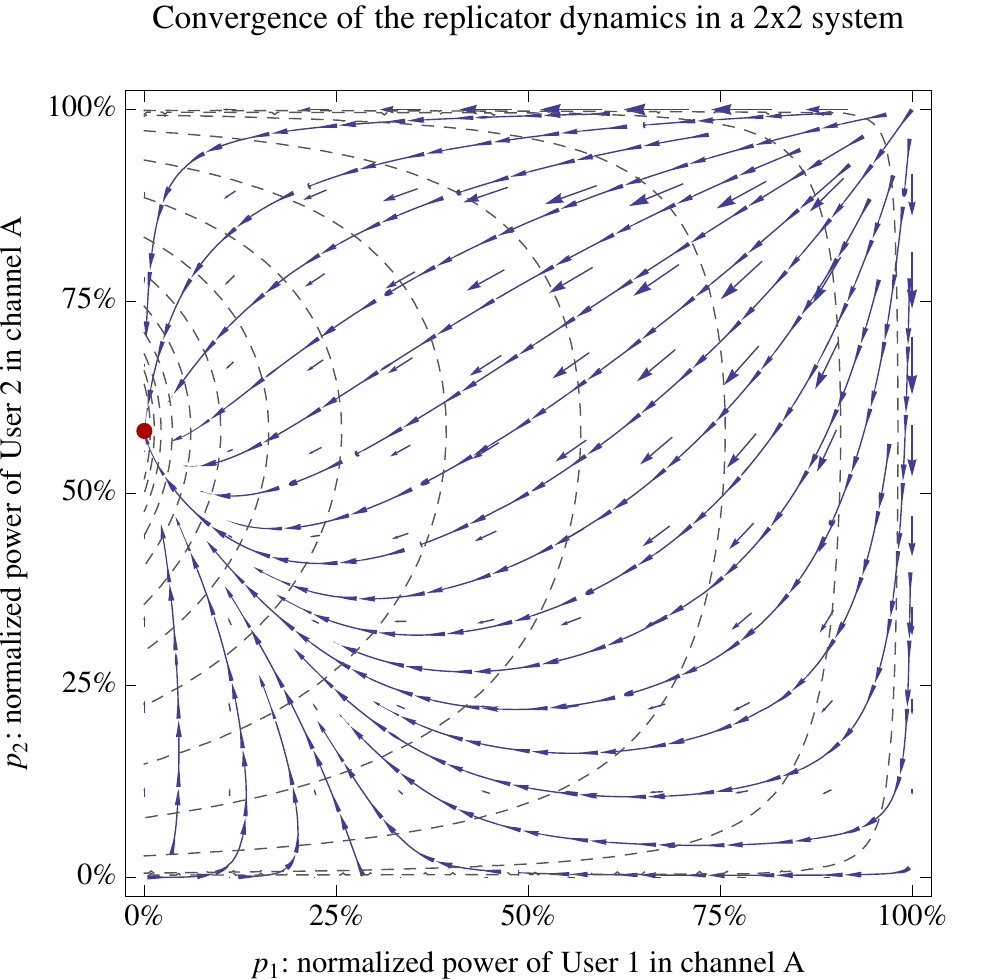}}
\hspace{30pt}
\subfigure[Spectral efficiency over time for different initial configurations.]{
\label{subfig:speed.power}
\includegraphics[width=.97\columnwidth]{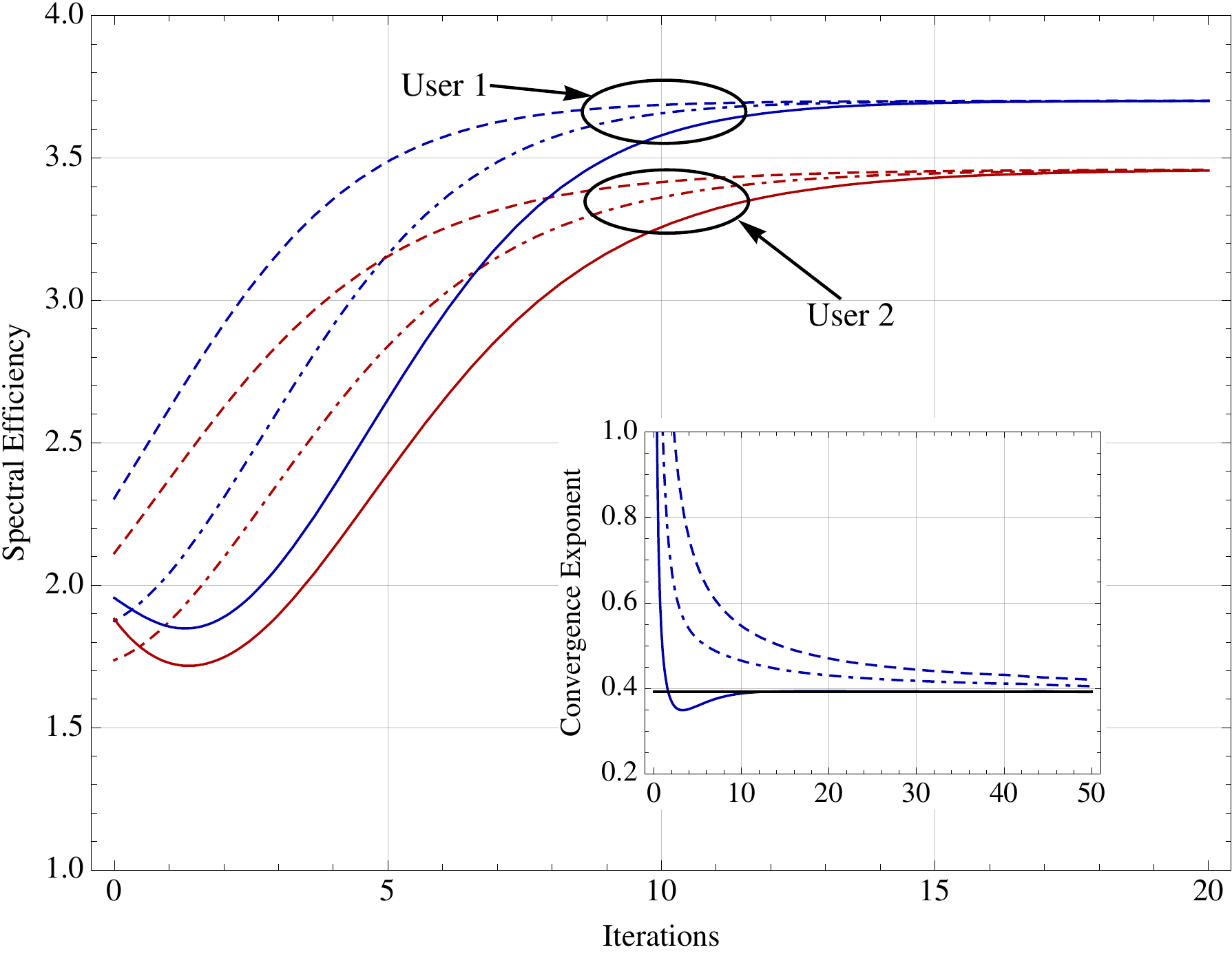}}
\caption{Convergence to equilibrium in a $2\times2$ game with static channels. The dashed contours in Fig.~\ref{subfig:portrait} are the level sets of the K-L divergence w.r.t. the game's equilibrium (red dot), while $p_{1}, p_{2}$ represent the normalized power allocation of each user. In Fig.~\ref{subfig:speed.power}, we plot the spectral efficiency $u_{k}$ of each user as a function of time for three randomly drawn initial configurations; as can be seen in the inlay, the equilibration rate $\lambda(t)$ coincides with the value predicted by Theorem \ref{thm:convergence} (solid black line).
\vspace{-15pt}}
\label{fig:convergence}
\end{figure*}

The first important property of (\ref{eq:RD}) is that its rest points satisfy the (waterfilling) condition $v_{k\alpha}(p) = v_{k\beta}(p)$ for all nodes $\alpha,\beta\in\supp(p_{k})$ to which user $k$ transmits with positive power. Hence, by the \ac{KKT} conditions of \cite{PBLD09}, we see that \emph{Nash equilibria of $\game$ are stationary in (\ref{eq:RD})}.

The converse of this statement is not true: every vertex of $\strat$ is stationary \emph{without} necessarily being a Nash equilibrium. Nonetheless, the game's (unique) Nash equilibrium is the {\em only} attracting state of the dynamics (see Appendix~\ref{apx:convspeed} for the proof):
\begin{theorem}
\label{thm:convergence}
Let $q\in\strat$ be the (a.s.) unique equilibrium of $\game$. Then, every interior solution orbit of the replicator dynamics (\ref{eq:RD}) converges to $q$; moreover, there exists $c>0$ such that
\begin{equation}
\label{eq:convspeed}
\dkl\left(q \midd p(t)\right) \leq \dkl\left(q\midd p(0)\right)\,e^{-ct}\,\,
\text{for all $t\geq0$,}
\end{equation}
where $\dkl$ is the Kullback-Leibler divergence. In other words, replicator trajectories converge to an $\eps$-neigh\-borhood of a Nash equilibrium in time $\bigoh(\log(1/\eps))$.
\end{theorem}

\setcounter{remark}{0}

\begin{remark}
To the best of our knowledge, the exponential convergence rate of Theorem \ref{thm:convergence} (see also Theorem \ref{thm:convspeed} in App.~\ref{apx:convspeed}  and Fig.~\ref{fig:convergence}) is the fastest known estimate for the replicator dynamics (see \cite{Sandholm11} for a review of the state of the art).
\end{remark}

\begin{remark}
The {\em Kullback-Leibler divergence} is defined as \cite{Sandholm11}:
\begin{equation}
\dkl(q\midd p)
= \insum_{k,\alpha:q_{k\alpha}>0} q_{k\alpha} \log\left({q_{k\alpha}}\big/{p_{k\alpha}}\right).
\end{equation}
Clearly, $\dkl(q_{k} \midd p_{k})$ is finite if and only if $p_{k}$ allocates positive power $p_{k\alpha}>0$ to all channels $\alpha\in\supp(q)$ which are present in $q_{k}$. Thus, in particular, Theorem \ref{thm:convergence} guarantees that uniform initial power allocations equilibrate exponentially quickly.
\end{remark}

\begin{remark}
In the evolutionary analysis of \cite{Sa01}, the fitness of a species (the number of descendants in the unit of time) might be nonlinear, but it is still a convex combination of each phenotype's fitness. In this special case, the dynamics of \cite{Sa01} are formally equivalent to (\ref{eq:RD}), and it is shown therein that their limit points are Nash equilibria. Theorem \ref{thm:convergence} (see also Theorem \ref{thm:convspeed}) extends this analysis by demonstrating that the replicator dynamics really \emph{do} converge to Nash equilibrium, and that the rate of this convergence is exponential.
\end{remark}

\begin{remark}
It was shown in \cite{SPB08-jsac} that iterative water-filling algorithms converge exponentially when the water-filling operator is a contraction. However, given that the sufficient conditions which guarantee the contraction property fail in the \ac{PMAC} case \cite{ValueTools11b}, the analysis of \cite{SPB08-jsac} does not apply here.
\end{remark}

Of course, the value of the exponent of (\ref{eq:convspeed}) is critical because it controls how fast users converge to equilibrium. Thus, if we consider the ``instantaneous'' convergence exponents:
\begin{equation}
\label{eq:exponent}
\lambda_{k}(t) \equiv -\frac{1}{t} \log \frac{\dkl(q_{k}\midd p_{k}(t))}{\dkl(q_{k}\midd p_{k}(0))},
\end{equation}
then Theorem \ref{thm:convergence} simply states that the total equilibration rate $\lambda(t) \equiv \min_{k}\{\lambda_{k}(t)\}$ is at least $c$. For the sake of simplicity, we will only present here an analytic expression for the value of $c$ for \emph{strict} equilibria (for the full analysis including non-strict equilibria, see Appendix \ref{apx:convspeed}). In this case, if user $k$ transmits with full power to channel $\alpha_{k}$ at equilibrium, we will have:
\begin{equation}
\label{eq:strictspeed}
\txs
c_{k} \equiv \liminf\nolimits_{t}\{\lambda_{k}(t)\}
=\gamma_{k}^{-1}\!\left(1-e^{-\gamma_{k}}\right) \Delta v_{k}
\text{ and }
c=\min\nolimits_{k}\{c_{k}\}
\end{equation}
where $\Delta v_{k} = \min\big\{v_{k,\alpha_{k}}(q) - v_{k\beta}(q):\beta\neq \alpha_{k}\big\}$ is $k$'s minimum deviation cost and $\gamma_{k} = \dkl(q\midd p(0))/P_{k}$. We thus obtain:

\begin{proposition}
\label{prop:powerconvergence}
If $q=\sum_{k}P_{k} e_{k,\alpha_{k}}$ is a strict equilibrium of the static game $\game$, the power of user $k$ on channel $\alpha_{k}$ grows as:
\begin{equation}
p_{k,\alpha_{k}}(t)\sim P_{k}\left(1-e^{-\Delta v_{k}\,t}\right).
\end{equation}
\end{proposition}

\begin{IEEEproof}
From (\ref{eq:strictspeed}), we have $c_{k}\to\Delta v_{k}$ as $\gamma_{k}\to 0$. However, since $\dkl(q_{k}\midd p_{k}(t))\to 0$ as $t\to\infty$ (Theorem \ref{thm:convergence}), we will have $\gamma_{k}\to0$ by definition, and our assertion follows.
\end{IEEEproof}


\subsection{Fading Channels}
\label{subsec:ergodic.convergence}

\subsubsection{Block-fading channels}

In this case, the replicator equation (\ref{eq:RD}) becomes non-deterministic because the coefficients $g_{k\alpha}$ evolve stochastically over time. To account for this, we will rewrite the replicator dynamics (\ref{eq:RD}) in discrete time as:
\begin{equation}
\label{eq:stochRD}
\Delta p_{k\alpha}(n+1) = \delta(n) p_{k\alpha}(n) \, \big[v_{k\alpha}(p(n),g(n)) - v_{k}(p(n),g(n))\big],
\end{equation}
where $\Delta p_{k\alpha}(n+1) \equiv p_{k\alpha}(n+1) - p_{k\alpha}(n)$ and the ``step'' $\delta(n)$ is a (possibly time-dependent) learning parameter.

For simplicity, we will concentrate here on the case where the temporal variations of the channels are uncorrelated. In this case, if we set $\overline v_{k\alpha} = \ex[v_{k\alpha}]$ and $\eta_{k\alpha} = v_{k\alpha} -\overline v_{k\alpha}$, we obtain:
\begin{flalign}
\label{eq:stochapproxRD}
p_{k\alpha}(n+1)
	&= p_{k\alpha}(n)
	+ \delta(n) p_{k\alpha}(n) \Big(\overline v_{k\alpha}(p(n)) - \overline v_{k}(p(n))\Big)\notag\\
	&+ \delta(n) p_{k\alpha}(n) \Big(\eta_{k\alpha}(p(n),g(n)) - \eta_{k}(p(n),g(n))\Big),
\end{flalign}
where the $\overline v_{k\alpha}$ are deterministic and the $\eta_{k\alpha}$ are zero-mean. In fact, if we interchange expectation and differentiation, we get:
\begin{equation}
\label{eq:ergderpayoff}
\overline v_{k\alpha}(p)
= \ex\left[\frac{\pd u_{k}}{\pd p_{k\alpha}}\right]
= \frac{\pd}{\pd p_{k\alpha}}\ex[u_{k}]
= \frac{\pd \overline u_{k}}{\pd p_{k\alpha}},
\end{equation}
so the mean utilities of (\ref{eq:stochapproxRD}) are the gradients of the ergodic rates (\ref{eq:ergpayoff}). Thus, if we remove the noise $\eta_{k\alpha}$ from (\ref{eq:stochapproxRD}), the general theory of \cite{Borkar08} shows that (\ref{eq:stochapproxRD}) will track the {\em mean-field equation}:
\begin{equation}
\label{eq:meanRD}
\frac{dp_{k\alpha}}{dt} = p_{k\alpha}(t)\,\big[\overline v_{k\alpha}(p(t)) - \overline v_{k}(p(t))\big],
\end{equation}
so the asymptotic properties of (\ref{eq:stochapproxRD}) will follow those of (\ref{eq:meanRD}).

We thus see that the asymptotic behavior of the replicator algorithm (\ref{eq:stochRD}) for block-fading channels is intimately linked to the ergodic game $\overline\game$. In itself, this is quite natural because the ergodic equilibrium of $\overline\game$ represents the only reasonable time-invariant equilibrial notion for the block-fading game $\game(t)$ with temporally uncorrelated channels (see also \cite{MM10}). More to the point, we have (see App.~\ref{apx:stochastic} for the proof):

\begin{theorem}
\label{thm:stochconvergence}
If the learning parameters $\delta(n)$ of (\ref{eq:stochRD}) satisfy $\sum_{n=1}^{\infty} \delta(n) = \infty$ and $\sum_{n=1}^{\infty} \delta^{2}(n) < \infty$, then the algorithm (\ref{eq:stochRD}) for temporally uncorrelated block-fading channels converges (a.s) to the (unique) Nash equilibrium of the ergodic game $\overline\game$.
\end{theorem}

\begin{remark*}
The most usual choice for the parameters $\delta(n)$ is $\delta(n) = 1/n$. These variable rates can be interpreted either as the actual time step of the algorithm, or as a discount that users apply to their updating scheme at every tick of a timer. This last interpretation is crucial for practical purposes because there are hard limits to how fast a device can update its policy. For constant $\delta(n)\equiv\delta$, the dynamics (\ref{eq:stochRD}) evolve faster, but convergence to equilibrium is in the distribution sense of \cite{Borkar08}.
\end{remark*}

\subsubsection{Fast-fading channels}

As we have already noted in Section \ref{subsec:fading.game}, the users' (ergodic) rates (\ref{eq:ergpayoff}) in the fast-fading regime depend on the channels' statistics, so instantaneous channel information obtained when updating their powers is of little use. Because of this, the system becomes effectively deterministic, so, similarly to the static case, users may base their learning on the (mean) replicator learning dynamics (\ref{eq:meanRD}) \textendash\ i.e. the discrete-time learning scheme (\ref{eq:stochRD}) without the noise term and with a constant step $\delta(n) = \delta$. We thus obtain (see App. \ref{apx:convspeed}):

\begin{theorem}
\label{thm:ergconvergence}
The mean dynamics (\ref{eq:meanRD}) converge to the (unique) Nash equilibrium of the ergodic game $\overline\game$, and this convergence is exponential: interior orbits $\eps$-equilibrate in time $\bigoh(\log(1/\eps))$.
\end{theorem}

\begin{remark*}
If the ergodic equilibrium is strict, the convergence exponent (\ref{eq:exponent}) of the mean dynamics (\ref{eq:meanRD}) is just (App.~\ref{apx:convspeed}):
\begin{equation}
\label{eq:strictstochspeed}
\txs
\overline c=\min\nolimits_{k}\left\{\overline c_{k}\right\}
\quad\text{with}\quad
\overline c_{k} =\gamma_{k}^{-1}\left(1-e^{-\gamma_{k}}\right) \Delta \overline v_{k},
\end{equation}
where $\gamma_{k} = \dkl(q\midd p(0))/P_{k}$ is as in (\ref{eq:strictspeed}), but now $\Delta \overline v_{k} = \min\big\{\overline v_{k,\alpha_{k}}(q) - \overline v_{k\beta}(q):\beta\neq \alpha_{k}\big\}$ is the minimum deviation cost of user $k$ for the \emph{mean} marginal utilities $\overline v_{k\alpha}$.
\end{remark*}


\section{Numerical Simulations}
\label{sec:numerics}

\begin{figure*}
\centering
\subfigure[Equilibrium \acl{SRE} for static channels.]{
\label{subfig:staticSRE}
\includegraphics[height=165pt,width=.9\columnwidth]{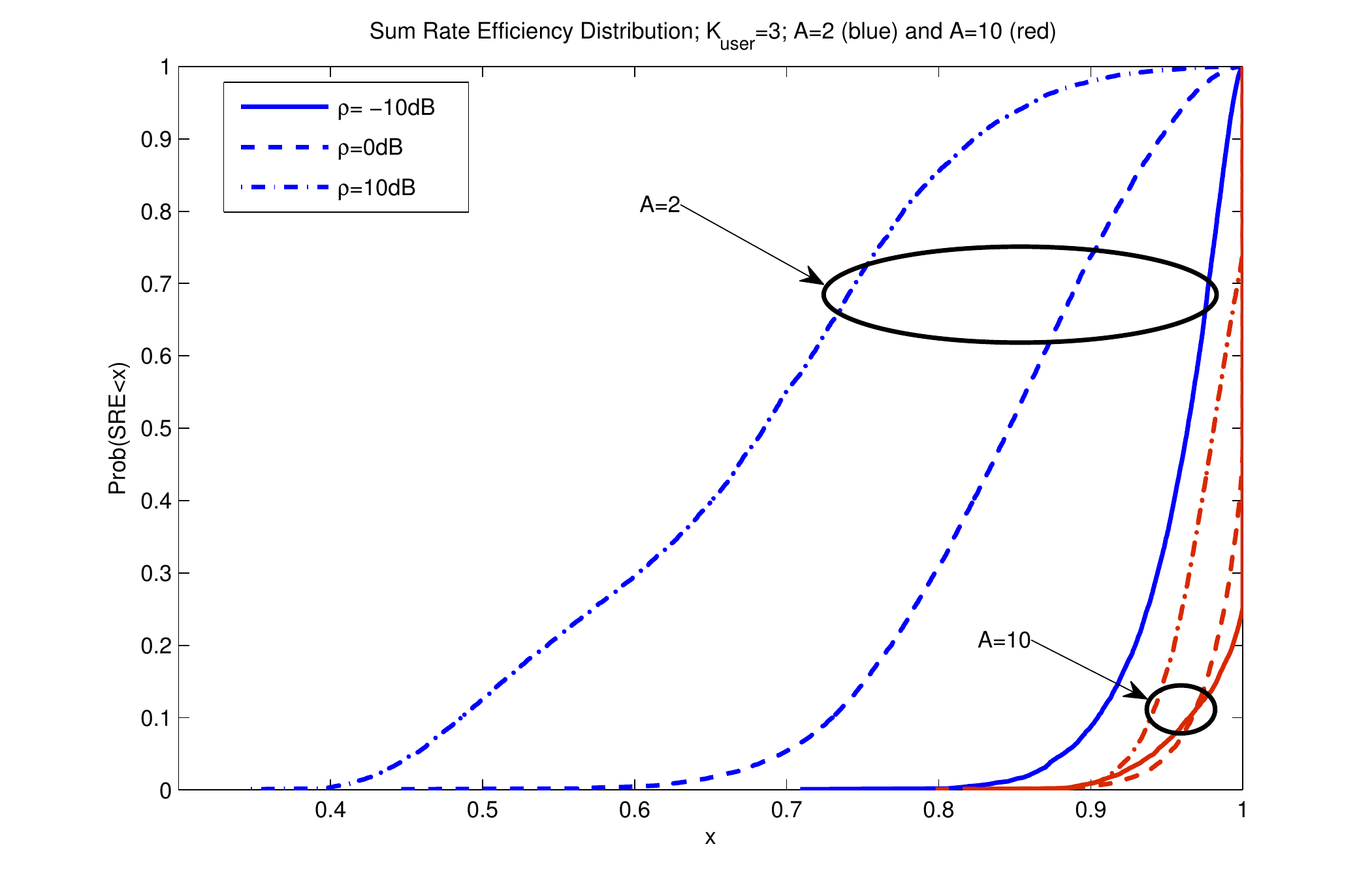}}
\qquad
\subfigure[Equilibrium \acl{SRE} for ergodic channels.]{
\label{subfig:ergodicSRE}
\includegraphics[height=165pt,width=.9\columnwidth]{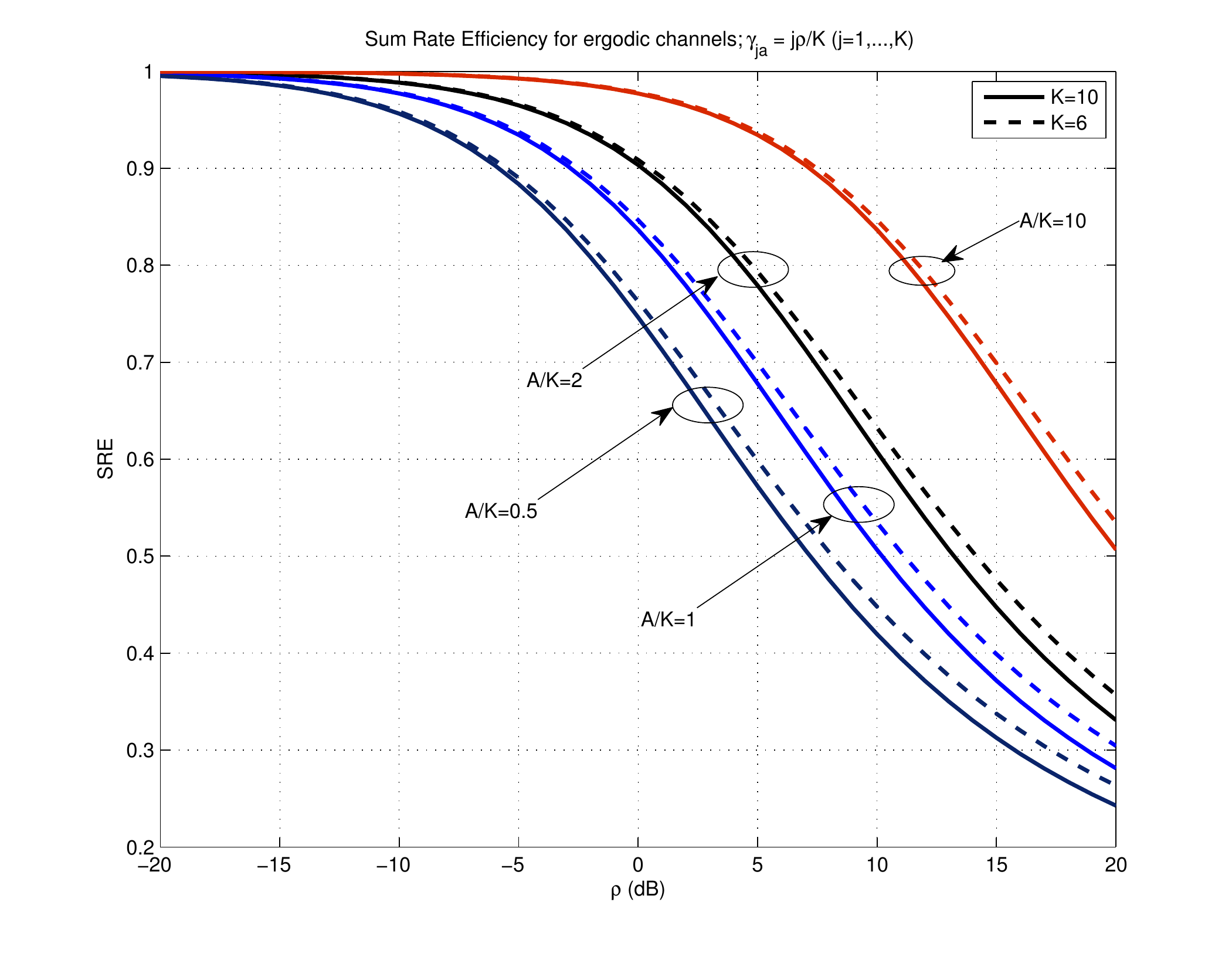}}
\caption{The \ac{CDF} of the equilibrium \acl{SRE} for static channels (Fig.~\ref{subfig:staticSRE}) and the equilibrium \ac{SRE} for ergodic Gaussian i.i.d. channels as a function of the thermal \ac{SNR} parameter $\rho = P_{\textit{max}}/\sigma^{2}$ (Fig.~\ref{subfig:ergodicSRE}) for different numbers of channels $A$ and users $K$ (all with similar maximum power constraints).}
\label{fig:SRE}
\end{figure*}

\begin{figure*}
\vspace{-10pt}
\centering
\subfigure[Sum-rate efficiency and equilibration over time for static channels.]{
\label{subfig:staticEQL}
\includegraphics[width=0.9575\columnwidth]{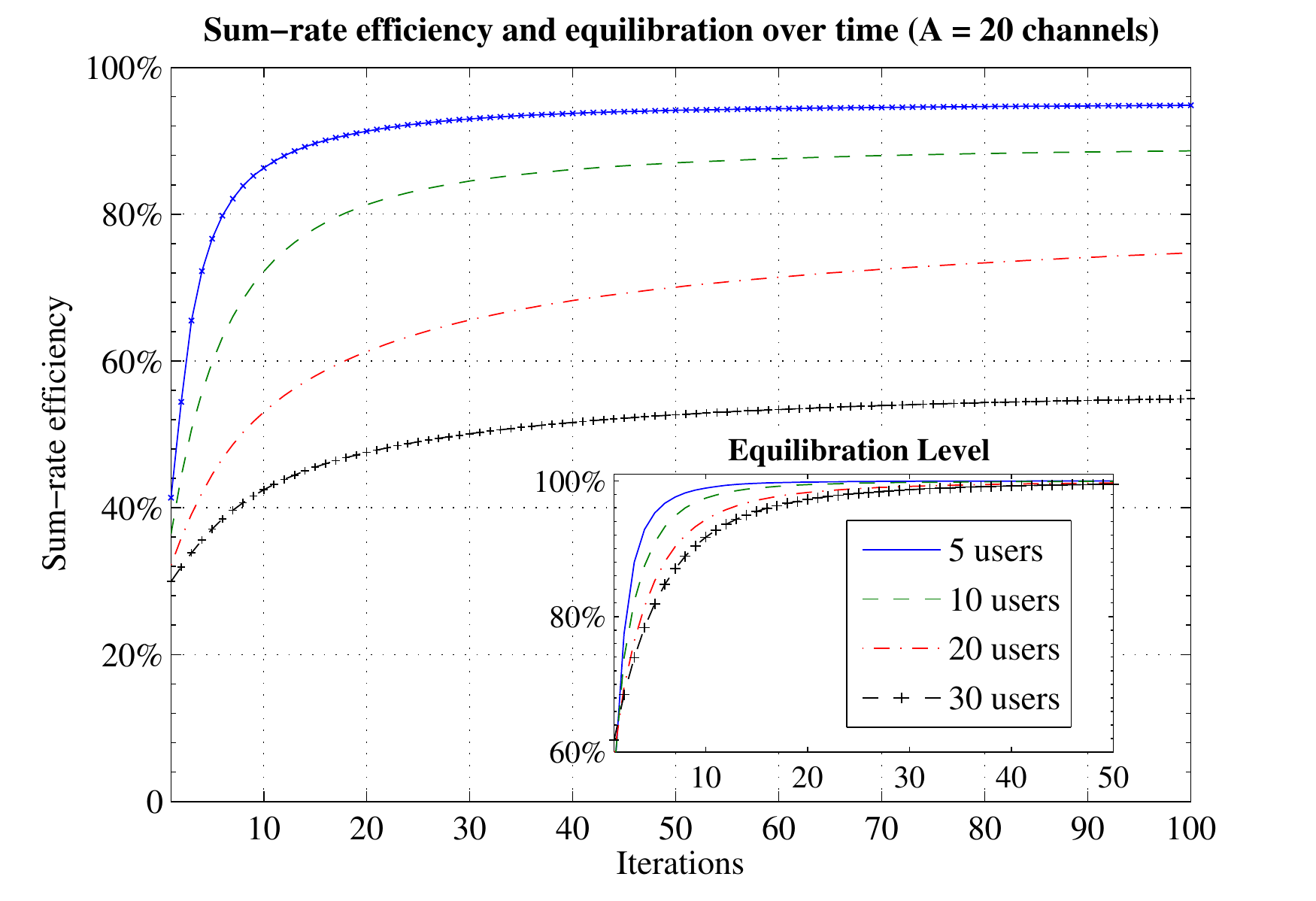}}
\subfigure[Long-term equilibration in block-fading channels.]{
\label{subfig:ergodicEQL}
\includegraphics[width=\columnwidth]{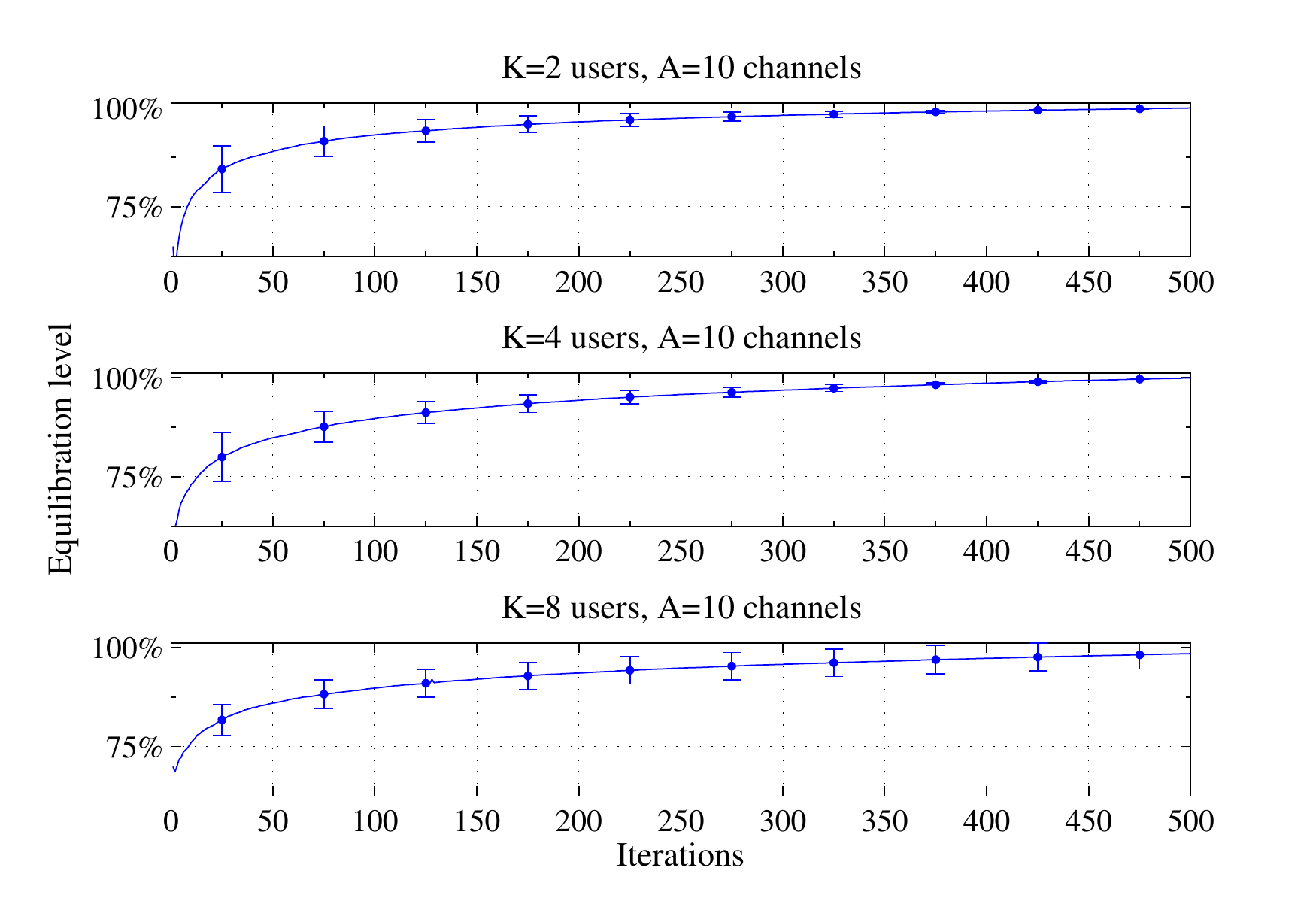}}
\caption{Equilibration (and its efficiency) for different numbers of users and channels. In static channels, the replicator dynamics equilibrate extremely fast, even for a large number of users; in temporally uncorrelated block-fading channels, the dynamics still converge, but slower \textendash\ due to the discounting $\delta(n) = 1/n$ in (\ref{eq:stochRD}).}
\label{fig:EQL}
\end{figure*}

\begin{figure*}
\centering
\subfigure{
\label{subfig:Jakes.follow.slow}
\includegraphics[width=\columnwidth]{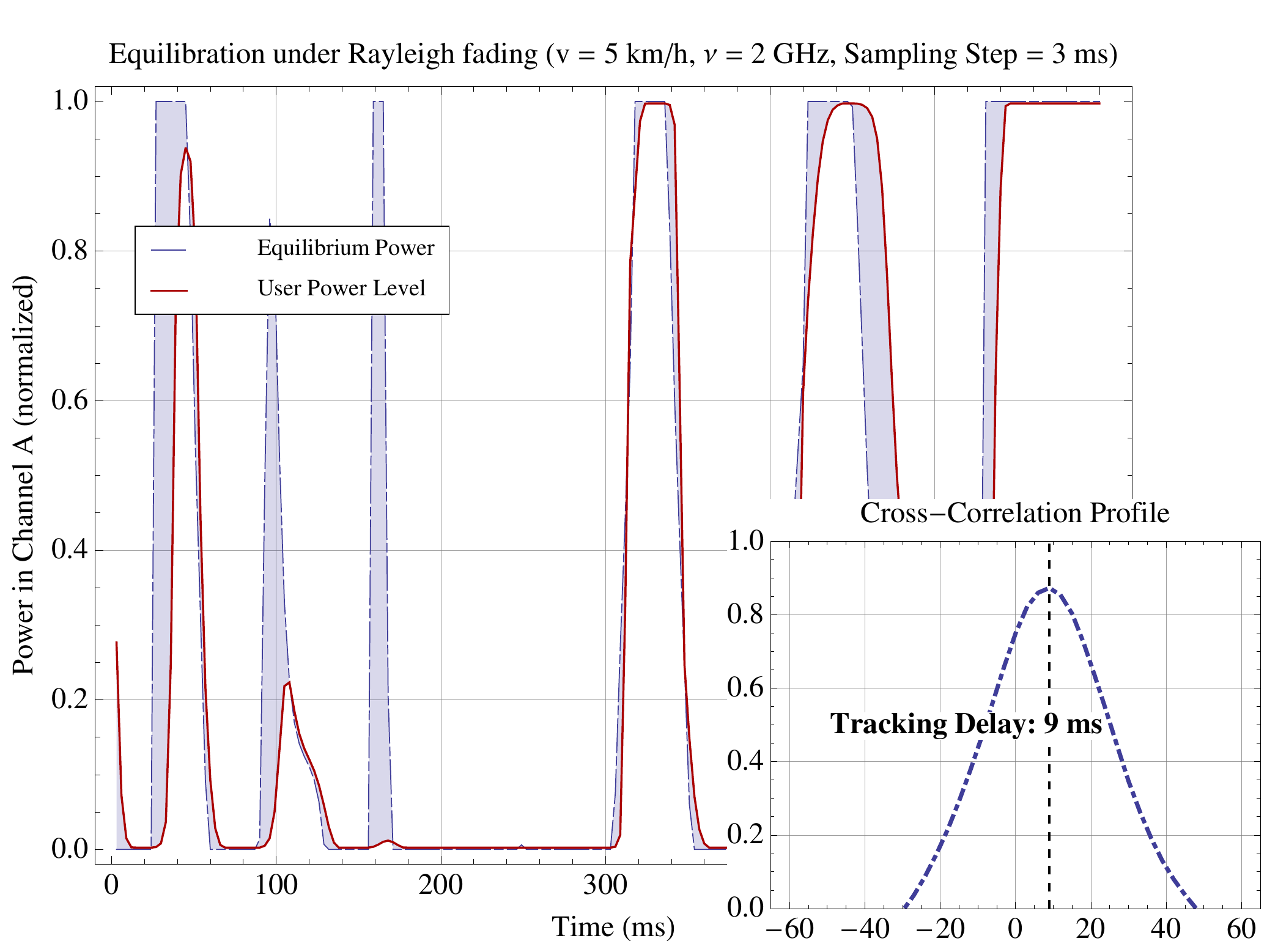}}
\hfill
\subfigure{
\label{subfig:Jakes.follow.fast}
\includegraphics[width=.95\columnwidth]{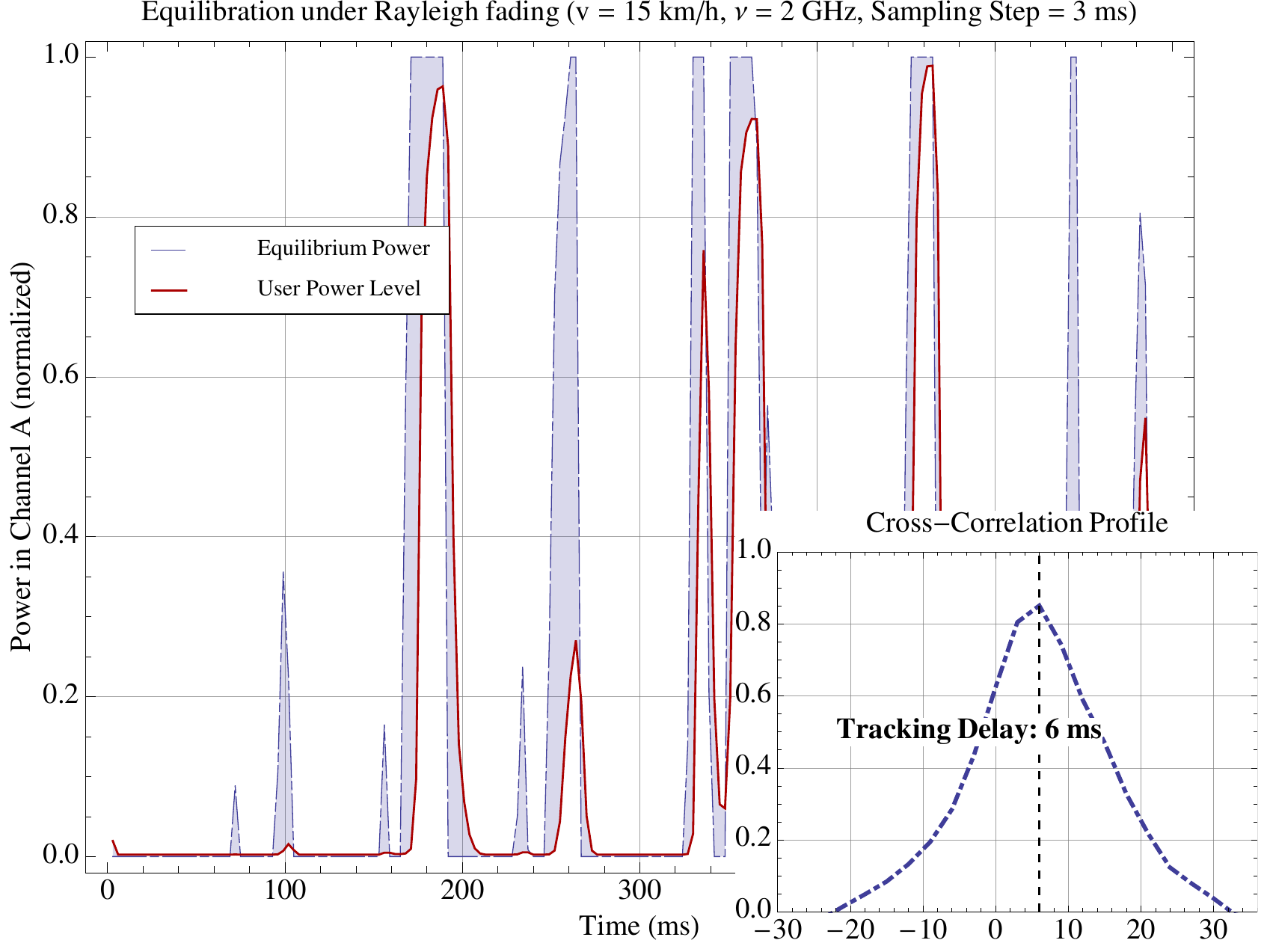}}
\caption{
The Nash power level (dashed blue) and the actual power level learned by a user (solid red) in a $2\times2$ Jakes-fading game. As can be seen by the cross-correlograms of the two time-series (inlays), users are away from equilibrium for only a very short amount of time (light blue shading): we observe a 9~ms tracking delay for user velocities in the 5~km/h range and 6~ms for 15~km/h, meaning that the replicator dynamics converge within 10-15\% of the system's coherence time (108~ms and 36~ms respectively).}
\label{fig:Jakes}
\end{figure*}

In this section, our aim is to validate our theoretical results by means of numerical simulations. We begin by introducing the \acfi{SRE} of a power profile $p$:
\begin{equation}
\label{eq:SRE}
\sre(p) = \frac{\sum_{k} u_{k}(p)}{C_{sum}},\quad
\text{(resp. $\dis \overline\sre(p) = \frac{\sum_{k} \overline u_{k}(p)}{\overline C_{sum}}$ for $\overline\game$)}
\end{equation}
i.e. the ratio of the sum of achievable rates in the power profile $p$ over the maximum achievable aggregate sum-rate under \ac{SIC} (which is the sum-capacity of the \ac{MAC}); interestingly, if $q$ is the game's (unique) equilibrium, then $C_{sum} = -\Phi(q)$ (and similarly for the ergodic case).

In Fig.~\ref{subfig:staticSRE}, we plot the \ac{SRE} at Nash equilibrium for randomly drawn static channels. While the equilibrium \ac{SRE} can deviate significantly from its maximum value (unity) for $A<K$, in the $A\geq K$ regime, the \ac{SRE} is typically close to 100\% (and, in fact, equal to 100\% with positive probability. We may attribute this to the fact that for $A\geq K$ there is a finite probability that the system's equilibrium is at a vertex of $\strat$ where each user is alone on a single channel, inducing optimal performance. Thus, with fair probability (close to $1/2$ based on our simulations) the complex \ac{SIC} scheme yields no performance benefit over the much simpler \ac{SUD} approach. Analogously, in Fig.~\ref{subfig:ergodicSRE}, we plot the \ac{SRE} at equilibrium for ergodic channels by using Proposition \ref{prop:gaussianpotential} to evaluate the maximum sum-rate under \ac{SIC}; in that case, while the \ac{SRE} is nearly optimal for small \ac{SNR}, it deviates strongly from its maximum value for larger \ac{SNR} values.

Furthermore, to test the equilibration rate of (\ref{eq:stochRD}) for different values of $K$ and $A$, we introduce the \ac{EQL}:
\begin{equation}
\label{eq:EQL}
\eql(p) = \Phi(p)\big/\Phi(q),\,\,
\text{(resp. $\dis \overline\eql(p) = \overline\Phi(p)\big/\overline\Phi(q)$ for $\overline{\game}$),}
\end{equation}
with $q$ being the equilibrium of $\game$ (resp.~$\overline\game$) \textendash\ so an \ac{EQL} of $1$ implies that the system has reached its equilibrium.

Beginning with the static case, in Fig.~\ref{subfig:staticEQL} we drew $N=50$ channel realizations and ran the discrete-time learning scheme (\ref{eq:stochRD}) with constant $g(n)$ and $\delta(n)$ for $A=20$ channels and $K=5,\,10,\,20$ and $30$ users. Then, by plotting the average \ac{SRE} and \ac{EQL} over time, we see that even for $30$ users, the system equilibrates within a few tens of iterations. On the other hand, for the ergodic block-fading scenario of Fig.~\ref{subfig:ergodicEQL}, we used $A=10$ channels and plotted the system's {\footnotesize $\overline\eql$} over time for $K=2, 4$ and $8$ users learning with $\delta(n) = 1/n$. As predicted by Theorem \ref{thm:stochconvergence}, the system converges to the game's ergodic equilibrium \textendash\ but slower due to the discounting $\delta(n) = 1/n$. Finally, for fast-fading users following the discrete-time version of (\ref{eq:meanRD}), the \ac{EQL} and \ac{SRE} plots were virtually identical to the static case (to be expected since both dynamical systems are deterministic), so they have been omitted for space considerations.

Finally, to test the convergence rate of the replicator dynamics in more realistic (non-ergodic) fading conditions that do not possess a long-term stationary equilibrium, we also simulated in Fig.~\ref{fig:Jakes} channels that follow the well-known Jakes model for Rayleigh fading \cite{Calcev++07}. Specifically, we considered a $2\times2$ game with user velocities $v=5$~km/h and $15$~km/h (Figs.~\ref{subfig:Jakes.follow.slow} and \ref{subfig:Jakes.follow.fast} respectively) transmitting at a carrier frequency of $\nu = 2\,\mathrm{GHz}$. We then ran the learning scheme (\ref{eq:stochRD}) with a constant update period of $\delta=3\,\mathrm{ms}$, and we plotted the (normalized) power level of a single user against the evolving equilibrium power level (calculated at each step based on the instantaneous channel coefficients). Then, to quantify how well the users follow the system's evolving equilibrium, we calculated the cross-correlation of the two processes and the users' \emph{tracking delay} (defined as the point of maximum cross-correlation).

Remarkably, Fig.~\ref{fig:Jakes} shows that the dynamics track the game's evolving equilibrium extremely closely. On average, users equilibrate within 9~ms for 5~km/h fading velocities, and within 6~ms for 15~km/h \textendash\ meaning that users converge within 10-15\% of the system's coherence time (108~ms and 36~ms respectively).

\begin{remark*}
For larger numbers of users (or channels), the results observed are similar; for instance, if the users in the previous game are increased to a few tens (we went up to $N=50$), their tracking delay becomes longer but never exceeds 30-35\% of the channel coherence time. However, due to space limitations, we opted to present here only the $2\times2$ case for simplicity.
\end{remark*}


\section{Conclusions and Future Directions}
\label{sec:conclusions}

In this paper, we studied the distributed power allocation problem for orthogonal uplink channels by introducing a game which admits a convex potential function. For both static and fading channels, we found that the associated game admits a unique Nash equilibrium and we showed that a simple distributed learning scheme based on the replicator dynamics converges to equilibrium from (almost) any initial condition. In fact, by proving a general result for convex potential games, we showed that the speed of this convergence is exponential: users converge to an $\eps$-neighborhood of an equilibrium in time which is at most of order $\bigoh(\log(1/\eps))$.

There is a number of important extensions of this work which demonstrate the strength of the replicator dynamics in continuous nonlinear games of this sort. First off, instead of the achievable rates $u_{k}$, one could consider energy-efficient metrics where users do not saturate their power constraints \textendash\ e.g., when the price of transmission power might restrain users from transmitting at maximum power. More importantly, these techniques can be extended even to non-orthogonal channel models such as the \ac{MIMO} \ac{MAC} case where the game's strategy space consists of all positive-definite precoding matrices with constrained trace. Because of this nonlinear structure, the form (\ref{eq:RD}) of the replicator dynamics no longer applies, but one can still write down a suitably modified {\em matrix-valued} replicator equation which allows users to converge to equilibrium.


\appendices


\section{Convergence Speed of the Replicator Dynamics}
\label{apx:convspeed}

Recall first that the \emph{solid tangent cone} to $\strat$ at $q$ is the set of rays starting at $q$ and in\-ter\-secting $\strat$ in at least one other point, i.e.: $\cone_{q}\strat \equiv \big\{z\in \R^{Q}: z_{k\alpha}\geq0 \text{ for all $\alpha\in\act_{k}$ with $q_{k\alpha}=0$}\big\}$.
With this in mind, we have the following generalization of convexity:
\begin{definition}
\label{def:starconvex}
A function $F\colon\strat\to\R$ will be called \emph{star-convex} w.r.t. $q\in\strat$ if $f(\theta)\equiv F(q+\theta z)$ is convex and increasing for all $z\in\cone_{q}\strat$ and for all $\theta>0$ s.t. $q+\theta z\in\strat$.
\end{definition}
Star-convex functions need not be convex, but strictly convex functions are star-convex w.r.t. their global minimum and weakly convex functions with a unique minimum are also star-convex \textendash\ in particular, both $\Phi$, $\overline\Phi$ are star-convex. For games with star-convex potentials, we then have:
\begin{theorem}
\label{thm:convspeed}
Let $\mathfrak{Q}\equiv\mathfrak{Q}\left(\play,\{\strat_{k}\}, \{\phi_{k}\}\right)$ be a game with a star-convex potential $F$. Then, the replicator dynamics (\ref{eq:RD}) for the marginal utilities $\phi_{k\alpha} = \frac{\pd \phi_{k}}{\pd p_{k\alpha}}$ converge to $q$ for any initial condition that starts at finite K-L divergence $h_{0}\equiv \dkl(q\midd p(0))$ from $q$. Moreover, there exists $c>0$ such that:
\begin{equation}
\label{eq:genconvspeed}
\dkl(q\midd p(t)) \leq h_{0}\,e^{-ct}\quad
\text{for all $t\geq0$}.
\end{equation}
\end{theorem}

Our proof strategy will be to establish an inequality of the form $\ddt H_{q}(p(t)) \leq -c H_{q}(p(t))$ and then employ Gr\"onwall's lemma. To that end, we first define the ``evolutionary index'':
\begin{equation}
\label{eq:evindex}
L_{q}(p) = - \txs\insum_{k,\alpha} \left(p_{k\alpha} - q_{k\alpha}\right) \phi_{k\alpha}(p),
\end{equation}
so named because $q$ is evolutionarily stable iff $L_{q}(p)>0$ near $q$. In fact, if we set $H_{q}(p) = \dkl(q\midd p)$, an easy calculation shows that $L_{q}$ is just (the negative of) the time derivative of $H_{q}$ w.r.t. the replicator dynamics (\ref{eq:RD}): $\ddt H_{q}(p(t)) = \insum_{k,\alpha} \frac{\pd H_{q}}{\pd p_{k\alpha}} \dot p_{k\alpha} =-L_{q}(p(t))$.

We will therefore begin by showing that $L_{q}(p)>0$ for all $p\in\strat\exclude\{q\}$, implying that $H_{q}$ is Lyapunov for the replicator dynamics (\ref{eq:RD}), and proving the convergence part of Theorem \ref{thm:convspeed}. Indeed, if we fix some $z\in\cone_{q}\strat$, (\ref{eq:evindex}) may be rewritten as $f'(\theta) = \insum_{k,\alpha} \left.\frac{\pd F}{\pd p_{k\alpha}}\right|_{q+\theta z} z_{k\alpha} = \theta^{-1} L_{q}(q+\theta z)$, for all $\theta>0$ such that $q+\theta z\in\strat$. Then, with $f(\theta)$ convex and increasing (Definition \ref{def:starconvex}), we obtain the estimate $L_{q}(p) = \theta f'(\theta) \geq f(\theta) - f(0) = F(p) - F(q)$, which shows that $L_{q}(p)>0$ for all $p\neq q$.

To prove the convergence time estimate (\ref{eq:genconvspeed}) we will need to show that $L_{q}$ grows linearly along directions which are not supported in $q$, and quadratically along those which {\em are} supported in $q$. To be specific, let $V_{q} = \{x\in\R^{\cardlinks}: \text{$x_{k\alpha}=0$ if $q_{k\alpha}=0$}\}$ be the subspace of directions of $\R^{\cardlinks}$, $\cardlinks = \sum_{k} A_{k}$, which are supported in $q$, and let $V_{q}^{\perp}$ be its orthocomplement in $\R^{\cardlinks}$. Then, by decomposing $z\in\R^{\cardlinks}$ as $z=z_{\parl}+z_{\perp}$ with $z_{\parl}\in V_{q}$ and $z_{\perp}\in V_{q}^{\perp}$, we define the seminorms $\|\cdot\|_{\parl}$ and $|\cdot|_{\perp}$ as:
\begin{align}
\label{eq:seminorms}
\txs
\|z\|_{\parl}^{2} \equiv \|z_{\parl}\|_{2}^{2} = \insum_{k,\alpha}^{\parl} z_{k\alpha}^{2},
&\quad
&
\txs
|z|_{\perp} \equiv \|z_{\perp}\|_{1} = \insum_{k,\alpha}^{\perp} |z_{k\alpha}|,
\end{align}
where the notation $\insum_{k,\alpha}^{\parl}$, $\insum_{k,\alpha}^{\perp}$ is shorthand for summing over the directions of $V_{q}$ and $V_{q}^{\perp}$ respectively.
We thus get:

\begin{lemma}
\label{lem:evindexestimate}
Let $F\colon\strat\to\R$ be star-convex w.r.t. $q\in\strat$. Then:
\begin{equation}
\label{eq:evindexestimate}
L_{q}(p) \geq F(p) - F(q) \geq m\,|p-q|_{\perp} + \tfrac{1}{2}r\,\|p-q\|_{\parl}^{2},
\end{equation}
where $m = \min_{k}\{\phi_{k\alpha}(q) - \phi_{k\mu}(q): q_{k\mu}=0,\, q_{k\alpha}>0\}$, and $r$ is the minimum of the Rayleigh quotient $\langle z,\mb M(q+z) z\rangle/\|z\|^{2}$ for the Hessian $\mb M(p) = \frac{\pd^{2}F}{\pd p_{k\alpha}\pd p_{\ell\beta}}$ of $F$, restricted over $\cone_{q}\strat$.
\end{lemma}

\begin{IEEEproof}
Since $q$ minimizes $F$, the \ac{KKT} conditions give $\phi_{k\alpha}(q) = -\left.\frac{\pd F}{\pd p_{k\alpha}}\right|_{q} = -\lambda_{k}$ for all $\alpha\in\act_{k}$ such that $q_{k\alpha}>0$ and $\phi_{k\alpha}(q)<-\lambda_{k}$ otherwise (where $\lambda_{k}$ denotes the complementary slackness Lagrange multiplier of $F$ over $\strat$). Thus, a first order Taylor estimate with Lagrange remainder readily yields:
\begin{equation}
\label{eq:fTaylor}
f(\theta) = f(0) + f'(0) \theta + \tfrac{1}{2} f''(\xi) \theta^{2}
\end{equation}
for some $\xi\in(0,\theta)$, so (\ref{eq:evindexestimate}) will follow once we properly estimate the linear and quadratic terms of (\ref{eq:fTaylor}).

As far as the linear term of (\ref{eq:fTaylor}) is concerned, we will have $f'(0)
= \insum_{k,\alpha} z_{k\alpha} \left.\frac{\pd F}{\pd p_{k\alpha}}\right|_{q}
= \insum_{k,\alpha}^{\parl} z_{k\alpha} \left.\frac{\pd F}{\pd p_{k\alpha}}\right|_{q}
+ \insum_{k,\alpha}^{\perp} z_{k\alpha} \left.\frac{\pd F}{\pd p_{k\alpha}}\right|_{q}
= \insum_{k,\alpha}^{\perp} z_{k\alpha} \left(\left.\frac{\pd F}{\pd p_{k\alpha}}\right|_{q} - \lambda_{k}\right)
\geq m\,|z|_{\perp}$, where the last equality holds because $\insum_{\alpha}^{\perp} z_{k\alpha} = - \insum_{\alpha}^{\parl} z_{k\alpha}$ (recall that $z\in\cone_{q}\strat$) and the last inequality is just the definition of $m$. Similarly, for any $\xi\in(0,\theta)$ and $z\in\cone_{q}\strat$, we get $f''(\xi)= \Big\langle z, \mb M(q+\xi z) z\Big\rangle=R_{q+\xi z}(\xi z)\,\|z\|^{2}$, where $R_{p}(w) = \big\langle w,\mb M(p) w\big\rangle$, $p\in\strat$, $w\in T_{p}\strat$, denotes the Rayleigh quotient of the Hessian $\mb M$ of $F$. Hence, if $r$ is the minimum of $R_{q+w}(w)$ over the set $B_{q} = \{w\in\cone_{q}\strat: q+w\in\strat\}$, we will also have $f''(\xi)\geq r \|z\|^{2}$, and (\ref{eq:evindexestimate}) follows by plugging the above into (\ref{eq:fTaylor}) and noting that $\|z\|\geq\|z\|_{\parl}$.
\end{IEEEproof}

Obtaining similar estimates for the relative entropy function $H_{q}$ is harder (after all, $H_{q}$ blows up near the boundary of $\strat$), so we will need two more auxiliary lemmas:
\begin{lemma}
\label{lem:uniqueroot}
For all $z\in\cone_{q}\strat\exclude\{0\}$ and for all $a>1$, the equation
\begin{equation}
\label{eq:uniqueroot}
\txs
H_{q}(q+\theta z) = a\,|z|_{\perp} \theta + \tfrac{1}{2} a \insum_{k,\beta}^{\parl} z_{k\beta}^{2}\big/ q_{k\beta} \, \theta^{2},
\end{equation}
admits a unique positive root $\theta_{a}\equiv\theta_{a}(z)$. Consequently:
\begin{equation}
\label{eq:directedestimate}
\txs
H_{q}(q+\theta z) \leq a\,|z|_{\perp} \theta + \tfrac{1}{2} a \insum_{k,\beta}^{\parl} z_{k\beta}^{2}\big/ q_{k\beta} \, \theta^{2}\,\,
\text{for all $\theta\leq\theta_{a}(z)$.}
\end{equation}
\end{lemma}

\begin{IEEEproof}
Let $h(\theta) \equiv H_{q}(q+\theta z)$ be the LHS of (\ref{eq:uniqueroot}), and denote its RHS by $ag(\theta)$. Then, if we set $w(\theta) = h(\theta) - ag(\theta)$, we readily obtain $w(0)=0$, $w'(0) = |z|_{\perp} (1-a)\leq 0$, and $w''(0) = \insum_{k,\beta}^{\parl} z_{k\beta}^{2}/q_{k\beta} (1-a)<0$, and the result follows by simple arguments relying on the mean value theorem.
\end{IEEEproof}

\begin{lemma}
\label{lem:entropyestimate}
Let $F\colon\strat\to\R$ be star-convex w.r.t. $q\in\strat$ and let $p(t)$ be a solution orbit of the replicator dynamics with initial relative entropy $h_{0} = H_{q}(p(0))$. Then, there exists $b>1$ s.t.:
\begin{equation}
\label{eq:entropyestimate}
\txs
H_{q}(p(t)) \leq b\,|p(t)-q|_{\perp} + \frac{b}{2q_{0}} \|p(t) - q\|_{\parl}^{2},\quad
\end{equation}
where $q_{0} = \min_{k,\alpha}\{q_{k\alpha}: q_{k\alpha}>0\}$.
\end{lemma}

\begin{IEEEproof}
Fix some $a>1$. Then, by Lemma \ref{lem:uniqueroot}, we know that (\ref{eq:uniqueroot}) admits a unique positive root $\theta_{a}(z)$, so let $h_{a}(z) = H_{q}(q+\theta_{a}(z) z)$ and set $h_{a} = \max\{h_{a}(z): z\in S_{q}\}$, where $S_{q} = \{z\in\cone_{q}\strat:\text{$z+q\in\strat$ but $q+(1+\eps)z\notin\strat$ for any $\eps>0$}\}$. Moreover, set $h_{c} = \max\{h_{0},h_{a}\}$, let $\theta_{c}(z)$ be the unique positive root of the equation $H_{q}(q+\theta_{c}(z) z) = h_{c}$, and define $b(z) = g(\theta_{c}(z))/h_{c}$ with $g(\theta) = |z|_{\perp} \theta + \tfrac{1}{2} \sum_{k,\beta}^{\parl} z_{k\beta}^{2}/q_{k\beta} \theta^{2}$ (as in the proof of Lemma \ref{lem:uniqueroot}). We will then have $b(z)\geq a$ since, otherwise, (\ref{eq:directedestimate}) would yield the contradiction $h_{c} = b(z) g(\theta_{c}(z)) < a g(\theta_{c}(z)) < h(\theta_{c}(z)) = h_{c}$.

With $b(z)>1$, a second application of Lemma \ref{lem:uniqueroot} yields $H_{q}(q+\theta z) \leq b(z) \left(|z|_{\perp} \theta + \tfrac{1}{2} \insum_{k,\beta}^{\parl} z_{k\beta}^{2}/q_{k\beta} \theta^{2}\right)$ for all $\theta\leq\theta_{c}(z)$. Thus, if we decompose $p(t)$ as $p(t) = q + \theta(t) z(t)$ with $\theta>0$ and $z(t)\in S_{q}$, we will have $\theta(t) \leq \theta_{c}(z(t))$; indeed, should this ever fail, we would have $H_{q}(p(t)) > b(z(t)) g(\theta(t)) > b(z(t)) g(\theta_{c}(t)) = h_{c} \geq h_{0}$ which contradicts the fact that $H_{q}$ is Lyapunov. Hence, with $\theta(t)\leq\theta_{c}(z(t))$ for all $t\geq0$, we get $H_{q}(p(t)) \leq b(z(t)) \left(|z(t)|_{\perp} \theta(t) + \tfrac{1}{2} \insum_{k,\beta}^{\parl} z_{k\beta}^{2}(t)/q_{k\beta} \theta^{2}(t)\right)$, and (\ref{eq:entropyestimate}) follows by taking $b = \max\{b(z):z\in S_{q}\}$.
\end{IEEEproof}

\begin{IEEEproof}[Proof of Theorem \ref{thm:convspeed}]
With notation as in Lemmas \ref{lem:evindexestimate} and \ref{lem:entropyestimate}, let $c=\min\{m/b,rq_{0}/b\}$. We then get $L_{q}(p(t)) \geq m\,|p(t)-q|_{\perp} + \tfrac{1}{2}r\,\|p(t)-q\|_{\parl}^{2} \geq c H_{q}(p(t))$ and Gr\"onwall's lemma yields $H_{q}(p(t)) \leq h_{0} e^{-ct}$. Since the \ac{KKT} inequalities for $F$ are strict along any direction of $\R^{\cardlinks}$ which is not supported in $q$, we will have $m>0$ and, consequently, $c>0$ as well.
\end{IEEEproof}

\begin{IEEEproof}[Proof of Theorems \ref{thm:convergence} and \ref{thm:ergconvergence}]
The potentials $\Phi$ and $\overline\Phi$ are star-convex, so both theorems follow from Theorem \ref{thm:convspeed}.
\end{IEEEproof}

All that remains is to calculate the value of $c$ when $q$ is strict. In that case, given that the intersection of $V_{q}$ with $\cone_{q}\strat$ is trivial, the quadratic term of (\ref{eq:evindexestimate}) can be ignored and we get $L_{q}(p) \geq \frac{1}{2} \insum_{k} \|p_{k}-q_{k}\|_{1} \Delta \phi_{k}$, where $\Delta \phi_{k} = \min_{\mu\neq\alpha_{k}}\{\phi_{k,\alpha_{k}}(q) - \phi_{k\mu}(q)\}>0$. As for (\ref{eq:entropyestimate}), we may decompose $p_{k}\in \strat_{k}\exclude\{q_{k}\}$ as $p_{k} = q_{k} + \theta_{k} z_{k}$ where $z_{k}\in \cone_{q_{k}}\strat_{k}$ has $z_{k,\alpha_{k}} = - P_{k}$. Thus, with $p_{k,\alpha_{k}} = P_{k}(1-\theta_{k})$, we readily obtain $H_{q}(p) = -\insum_{k} P_{k} \log(1-\theta_{k})$. Now, let $\theta_{k}^{*}$ be defined by the equation $h_{0} = H_{q_{k}}(q_{k}+\theta_{k}z_{k})$, i.e., $\theta_{k}^{*} = 1 -\exp(-h_{0}/P_{k})$, implying that $-P_{k}\log(1-\theta_{k}) \leq h_{0} \theta_{k}/\theta_{k}^{*}$ iff $0\leq\theta_{k}\leq\theta_{k}^{*}$ (because of convexity). We then claim that $H_{q}(p(t)) = -\insum_{k}P_{k} \log(1-\theta_{k}(t)) \leq h_{0} \insum_{k} \theta_{k}(t)/\theta_{k}^{*}$, where $\theta_{k}(t)$ is defined via the decomposition $p_{k}(t) = q_{k} + \theta_{k}(t) z_{k}(t)$.

However, if $\theta_{k}(t)>\theta_{k}^{*}$ for some $t\geq0$, then we would have $H_{q_{k}}(p_{k}(t)) > h_{0}$, and, hence, $H_{q}(p(t)) > H_{q}(p(0))$ as well, a contradiction \textendash\ recall that $H_{q}(p(t))$ is decreasing. Thus, combining all the above, we only need pick $c$ such that $P_{k}\,\Delta\phi_{k}\geq c h_{0}/\theta_{k}^{*}$, and the sharpest such choice is:
\begin{equation}
\label{eq:strictexponent}
c=\min\nolimits_{k}\left\{P_{k}/h_{0}\left(1-e^{-h_{0}/P_{k}}\right)\Delta \phi_{k}\right\}.
\end{equation}

\section{Stochastic Approximation of the Replicator Dynamics}
\label{apx:stochastic}

\begin{proof}[Proof of Theorem \ref{thm:stochconvergence}]
Note first that $\strat$ is invariant under the dynamics (\ref{eq:stochapproxRD}) if the $\delta(n)$ are chosen small enough. To see this, we will restrict ourselves w.l.o.g. to a game with one user and two choices, A and B (the general argument being similar). Thus, if we let $p_{A}(n) \equiv p_{1,A}(n)$ be the power that the user sends to channel A at the $n$-th iteration of the dynamics, we must find $\delta(n)$ such that $0\leq p(n) \leq 1$ for all $n\geq 0$ and all possible $g_{A,B}(n)\geq0$. So, assuming this holds for some $n\geq0$, we get:
\begin{multline}
\label{eq:diffRD}
p_{A}(n+1) - p_{A}(n)
= \delta(n) \,p_{A}(n) (1-p_{A}(n))\\
\times\left(
\frac{g_{A}(n)}{\sigma_{A}^{2} + g_{A}(n) p_{A}(n)} - \frac{g_{B}(n)}{\sigma_{B}^{2} + g_{B}(n) (1-p_{A}(n))}
\right).
\end{multline}
The first term of the LHS of (\ref{eq:diffRD}) is positive and the second is uniformly bounded, say by $M$, so $\delta(n)\leq M$ yields $p_{A}(n+1)\geq0$. The complementary inequality $p_{A}(n+1)\leq 1$ then follows similarly, so, with $p(n)\in\strat$ for all $n$, our theorem follows from Theorem~2 and Corollary~4 in Chap.~2 of \cite{Borkar08}.
\end{proof}


\bibliographystyle{ieeetran}
\footnotesize
\bibliography{IEEEabrv,Bibliography}


\balance

\begin{IEEEbiography}
[{\includegraphics[width=1.0in,height=1.25in,clip,keepaspectratio]{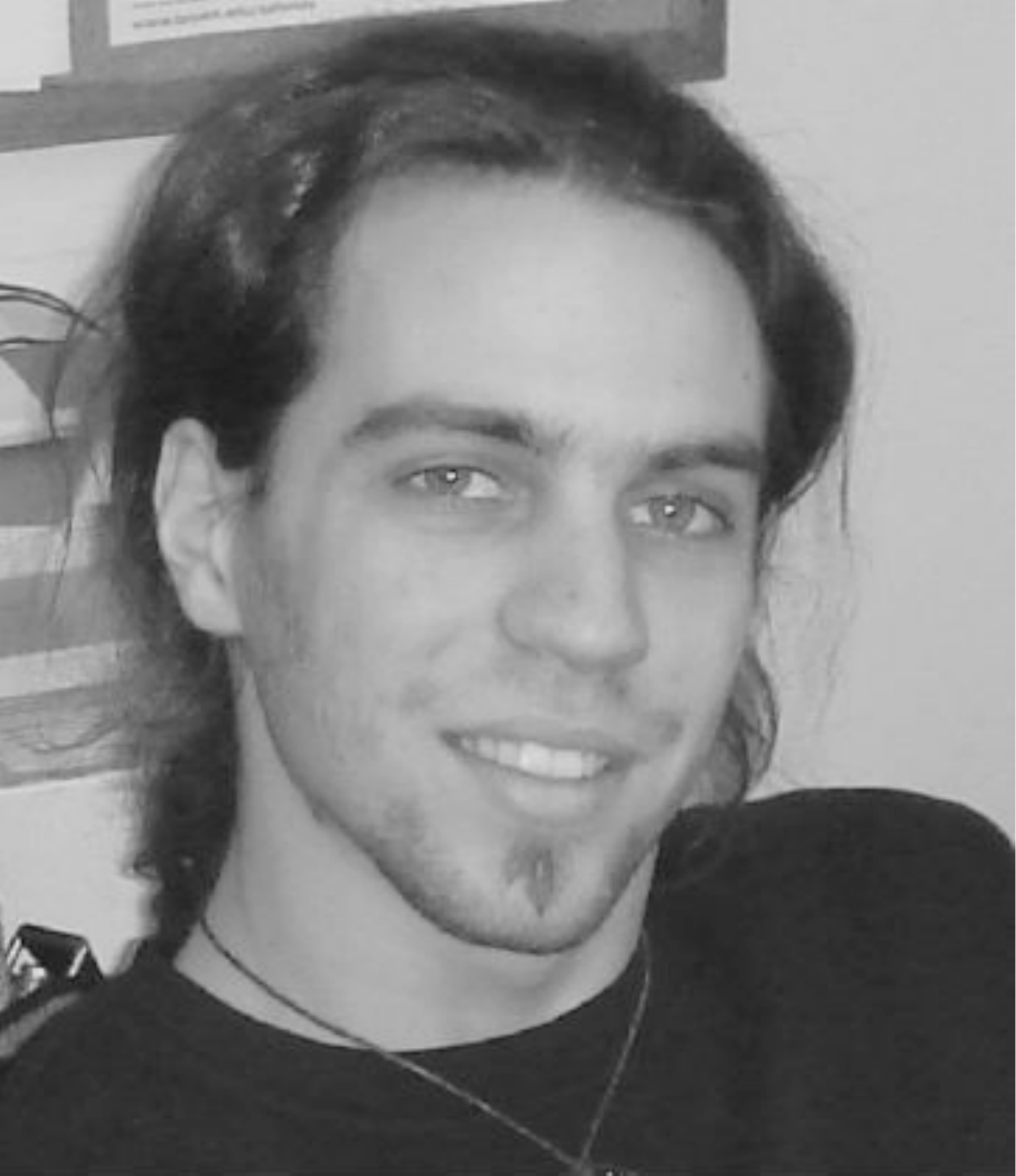}}]
{Panayotis Mertikopoulos}
received the B.S. degree in physics with a major in astrophysics from the University of Athens in 2003, the M.Sc. and M.Phil. degrees in mathematics from Brown University in 2005 and 2006 respectively, and the Ph.D. degree in physics from the University of Athens in 2010. In 2010-2011 he was with the Economics Department of \'Ecole Polytechnique. He is currently a CNRS Researcher at the Laboratoire d'Informatique de Grenoble, France.

Dr. Mertikopoulos has been an Embeirikeion Foundation Fellow since 2003. His research interests lie in dynamical systems, stochastic analysis, game theory, and their applications to networks.
\end{IEEEbiography}

\begin{IEEEbiography}
[{\includegraphics[width=1.0in,height=1.25in,clip,keepaspectratio]{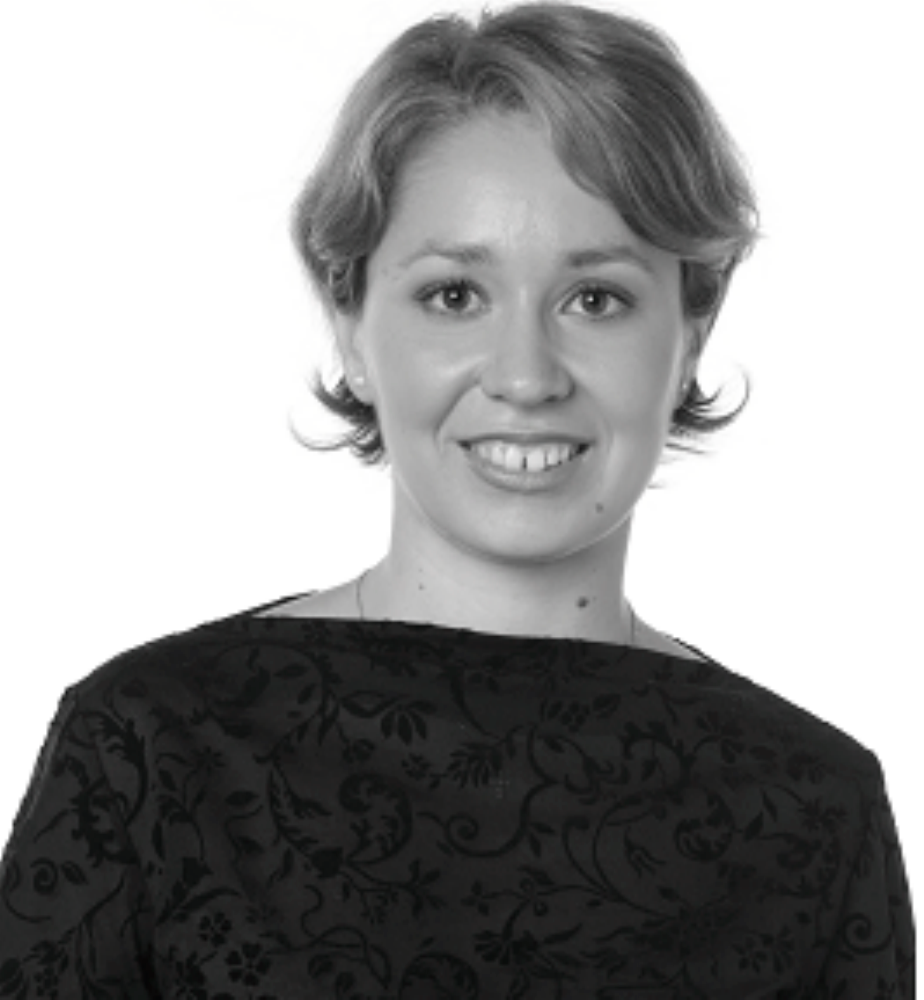}}]
{Elena V.~Belmega}
received the B.S. degree from the University Politehnica of Bucharest, Rom\^ania, in 2007. She obtained the M.Sc. and Ph.D. degrees, both from the Universit\'e Paris-Sud 11, Paris, France, in 2007 and 2010 respectively. In 2010-2011, Dr. Belmega was a post-doctoral researcher in a joint project between the Alcatel-Lucent Chair on Flexible Radio in Sup\'elec and Princeton University.  She is currently an assistant professor at ETIS/ENSEA \textendash\ Universit\'e de Cergy-Pontoise \textendash\ CNRS, Cergy-Pontoise, France.

Dr.~Belmega was one of the recipients  of the 2009 L'Or\'eal \textendash\ UNESCO \textendash\ French Academy of Sciences Fellowship: "For young women doctoral candidates in science".
\end{IEEEbiography}

\begin{IEEEbiography}
[{\includegraphics[width=1.0in,height=1.25in,clip,keepaspectratio]{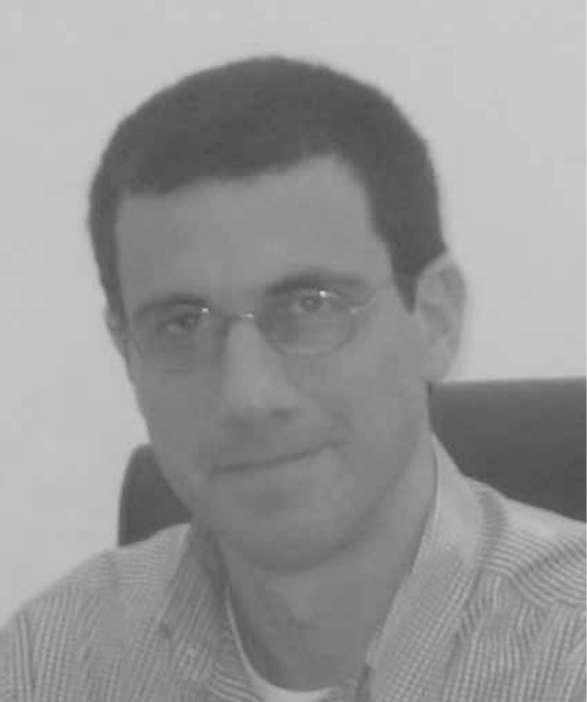}}]
{Aris Moustakas}(SM'04)
received the B.S. degree in physics from Caltech in 1990 and the M.S. and Ph.D. degrees in theoretical condensed matter physics from Harvard University in 1992 and 1996, respectively. He joined Bell Labs, Lucent Technologies, NJ, in 1998, first in the Physical Sciences Division and then also in the Wireless Advanced Technology Laboratory. He is currently an assistant professor at the Physics Department of the University of Athens, Greece.

Dr. Moustakas is serving as Associate Editor for the \textsc{IEEE Transactions on Information Theory}. His main research interests lie in the areas of multiple antenna systems and the applications of game theory and statistical physics to the theory of communications and networks.
\end{IEEEbiography}

\begin{IEEEbiography}
[{\includegraphics[width=1.0in,height=1.25in,clip,keepaspectratio]{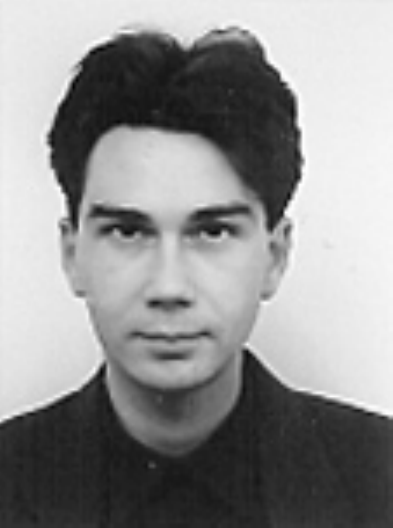}}]
{Samson Lasaulce}
received the B.S. degree in Applied Physics from \'Ecole Normale Sup\'erieure (Cachan) and the M.Sc. and Ph.D. degrees in Signal Processing from \'Ecole Nationale Sup\'erieure des T\'el\'ecommunications, Paris, France. He was with Motorola Labs from 1999 to 2001 and with France T\'el\'ecom R\&D in 2002 and 2003. Since 2004, he is with CNRS and Sup\'elec as a Senior Researcher and \'Ecole Polytechnique as professor.

Dr. Lasaulce is the recipient of the best student paper awards in \textsc{ValueTools '07} and \textsc{CrownCom '09}, and the best paper award of \textsc{NetGCoop '11}. He organized \textsc{GameComm '09} and \textsc{WNC3 '08}, and was the general chair of \textsc{ValueTools '11}. Dr. Lasaulce is serving as Associate Editor for the \textsc{IEEE Transactions on Signal Processing}. His research interests lie in communications, signal processing and information theory with a focus on game theory for wireless communications.
\end{IEEEbiography}

\end{document}